\documentclass[
10pt, % Main document font size
a4paper, % Paper type, use 'letterpaper' for US Letter paper
oneside, % One page layout (no page indentation)
headinclude,footinclude, % Extra spacing for the header and footer
BCOR5mm, % Binding correction
]{scrartcl}

\usepackage[
nochapters, % Turn off chapters since this is an article        
beramono, % Use the Bera Mono font for monospaced text (\texttt)
eulermath,% Use the Euler font for mathematics
pdfspacing, % Makes use of pdftex’ letter spacing capabilities via the microtype package
dottedtoc % Dotted lines leading to the page numbers in the table of contents
]{classicthesis} % The layout is based on the Classic Thesis style

\usepackage{arsclassica} % Modifies the Classic Thesis package

\usepackage[T1]{fontenc} % Use 8-bit encoding that has 256 glyphs

\usepackage[utf8]{inputenc} % Required for including letters with accents

\usepackage{graphicx} % Required for including images
\graphicspath{{Figures/}} % Set the default folder for images
\usepackage{paralist} % make inline lists
\usepackage{enumitem} % Required for manipulating the whitespace between and within lists

\usepackage{lipsum} % Used for inserting dummy 'Lorem ipsum' text into the template

\usepackage{subfig} % Required for creating figures with multiple parts (subfigures)

\usepackage{amsmath,amssymb,amsthm} % For including math equations, theorems, symbols, etc

\usepackage{varioref} % More descriptive referencing

\theoremstyle{definition} % Define theorem styles here based on the definition style (used for definitions and examples)

\theoremstyle{plain} % Define theorem styles here based on the plain style (used for theorems, lemmas, propositions)

\hypersetup{
colorlinks=true, breaklinks=true, bookmarks=true,bookmarksnumbered,
urlcolor=webbrown, linkcolor=RoyalBlue, citecolor=webgreen, % Link colors
pdftitle={}, % PDF title
pdfauthor={\textcopyright}, % PDF Author
pdfsubject={}, % PDF Subject
pdfkeywords={}, % PDF Keywords
pdfcreator={pdfLaTeX}, % PDF Creator
pdfproducer={LaTeX with hyperref and ClassicThesis} % PDF producer
} % Include the structure.tex file which specified the document structure and layout

\usepackage{lipsum} % Required to insert dummy text. To be removed otherwise
\usepackage{amsmath,amssymb,amsthm,amsfonts,scalerel}
\usepackage{color}
\usepackage{tikz}
\usepackage{pgfplots}
\usepackage{relsize}
\usepackage{float}
\usetikzlibrary{shapes,arrows}
\usepackage{bm,times}
\usepackage{verbatim}
\usetikzlibrary{calc,shadows,angles, arrows.meta, quotes}
\usepackage{enumitem}
\usepackage{multirow}
\usepackage{eucal}         % font produces calligraphic letters using mathcal e.g curly H for Hilbert space

\usepackage{colortbl}
\usepackage{siunitx}
\usepackage[title]{appendix}

\definecolor{kugray5}{RGB}{224,224,224}

\numberwithin{equation}{section}
\DeclareMathAlphabet{\pazocal}{OMS}{zplm}{m}{n}

\newtheorem{mydef}{Definition}[section]
\newtheorem{mytheo}{Theorem}[section]
\newtheorem{remark}{Remark}[section]
\newtheorem{corollary}{Corollary}[section]
\newtheorem{myprop}{Proposition}[section]

\setlist[enumerate,1]{%
	label=\arabic*.,
}

\newlist{inlinelist}{enumerate*}{1}
\setlist*[inlinelist,1]{%
	label=(\roman*),
}

\title{\normalfont\spacedallcaps{Unsmoothed Particle Hydrodynamics }\\ \large{exact resolution of identity method}} % The article title

\author{\spacedlowsmallcaps{Kalale Chola\textsuperscript{1} }} % The article author(s) - author affiliations need to be specified in the AUTHOR AFFILIATIONS block

\date{\today} % An optional date to appear under the author(s)

\begin{document}

%----------------------------------------------------------------------------------------
%	HEADERS
%----------------------------------------------------------------------------------------

\renewcommand{\sectionmark}[1]{\markright{\spacedlowsmallcaps{#1}}} % The header for all pages (oneside) or for even pages (twoside)
\lehead{\mbox{\llap{\small\thepage\kern1em\color{halfgray} \vline}\color{halfgray}\hspace{0.5em}\rightmark\hfil}} % The header style

\pagestyle{scrheadings} % Enable the headers specified in this block

%----------------------------------------------------------------------------------------
%	TABLE OF CONTENTS & LISTS OF FIGURES AND TABLES
%----------------------------------------------------------------------------------------

\maketitle % Print the title/author/date block

\setcounter{tocdepth}{2} % Set the depth of the table of contents to show sections and subsections only

%\tableofcontents % Print the table of contents

%\listoffigures % Print the list of figures

%\listoftables % Print the list of tables

%----------------------------------------------------------------------------------------
%	ABSTRACT
%----------------------------------------------------------------------------------------

\section*{Abstract} % This section will not appear in the table of contents due to the star (\section*)
\paragraph*{}The aim of this paper is to introduce a new computational fluid dynamics method to be called unsmoothed particle hydrodynamics SPH$-i$ which makes few assumptions and makes no assumption beyond the Navier-Stokes equations. The most attractive feature when compared with standard smoothed particle hydrodynamics (SPH) is that no explicit turbulence modeling is required. Furthermore, despite being a high order model, it retains the same, simple structure as standard SPH. In this sense SPH$-i$ is a coarse-grained direct numerical simulation approach.
   
In SPH, due the scale-dependence resulting from the convolution operator, all modes below the kernel cut-off length, are filtered out leading to loss of information. However, we conjecture that the SPH field, theoretically,  still contains enough information so that the SPH$-i$ field is a restored form of the original underlying continuum field. Since two filters are required, a rigorous technique for constructing compatible convolution and deconvolution filters is presented. The SPH$-i$ model is relatively easy to implement.

%\lipsum[1] % Dummy text

%----------------------------------------------------------------------------------------
%	AUTHOR AFFILIATIONS
%----------------------------------------------------------------------------------------

%{\let\thefootnote\relax\footnotetext{* \textit{Shintake Unit, OIST Graduate University, Okinawa, Japan}}}

{\let\thefootnote\relax\footnotetext{\textsuperscript{1} \textit{Shintake Unit, OIST Graduate University, Okinawa, Japan}}}

%----------------------------------------------------------------------------------------

\newpage % Start the article content on the second page, remove this if you have a longer abstract that goes onto the second page

%----------------------------------------------------------------------------------------
%	INTRODUCTION
%----------------------------------------------------------------------------------------

\section{Introduction}\label{sect:uph}
The smoothed particle hydrodynamics (SPH) method is a meshless particle method, originally developed for astrophysical problems and independently developed by Lucy \cite{Lucy1977} and Gingold and Monaghan \cite{Gingold1977}. The method has been successfully applied to mesoscopic and macroscopic flow problems \cite{Monaghan2005,Hu2006,Espanol2003,Ellero2018}. 

One particular area of interest in SPH applications is free surface flows which are generally turbulent. For high fidelity of the computed solutions, direct numerical simulation (DNS) is the best choice. However, due to the high demand of computational resources, DNS for free surface flows is generally not feasible. An alternative to DNS is large eddy simulation LES as it is computationally more efficient. 

An important similarity between SPH and explicit LES exists; they are both based upon the same integral interpolant. Therefore, it is possible to derive an SPH model that is consistent with explicit LES. This way, turbulence models developed in the LES context can be easily adopted in SPH. The trick is that since the SPH particle is by definition a "smoothed particle", it must move with the smoothed velocity. A rigorous derivation of this version of SPH has been presented in \cite{Cholas2018}.  

The problem of turbulence modeling is a relatively new topic in SPH. In 2002 Monaghan \cite{Monaghan2002} was able to show the similarity between his XSPH model and the Lagrangian-Averaged Navier Stokes LANS-$\alpha$ proposed in \cite{Holm1999,Holm2002}. This approach initially proved to yield promising results but was found to be computationally inefficient. It has however, undergone some refinements over the years \cite{Monaghan2009,Monaghan2011} with some success. Another SPH turbulence model based on the LANS-$\alpha$ model was proposed in \cite{Hu2015} and they demonstrated that their model was able to reproduce both the the inverse energy cascade and the direct enstrophy cascade of the kinetic energy spectrum. Dalrymple and Rogers \cite{Dalrymple2006} introduced a sub-particle scaling technique using the LES approach. The basic methodology is that the governing equations are averaged over a length scale specified by the filter and is comparable to the size of a fluid particle. This means that large scale motion (large eddies) will be fully resolved by solving the averaged equations. The effect of the unresolved small length scales (small eddies) on the large scale motion is contained in the sub-particle stress tensor which has to be modeled. Since the link between SPH and LES was not clearly demonstrated in this \cite{Dalrymple2006} and other work \cite{Violeau2007}, we recently developed a coherent SPH model that is consistent with explicit LES \cite{Cholas2018}.            

As an extension of our previous work \cite{Cholas2018}, in this paper a coarse-grained direct numerical simulation particle system that involves both filtering and de-filtering. The sub-particle filters appearing after the filtering process are de-filtered out, hence the resulting model requires no turbulence modeling.
  
The discussion in this paper will proceed as follows. First the filtering integral transform (FIT) and its associated de-filtering integral transform (DIT) are introduced. We then discuss the link between the FIT and DIT and further propose the procedure for constructing the deconvolution filter necessary for the inverse transform. Finally, a coarse-grained particle method consistent with implicit LES and based on spatial de-filtering, using the DIT, of the filtered CNSEs is derived. 

\section{Filtering Process}
The convolution or filtering problem can be stated formally as:
Given the continuum field\{$\rho(\mathbf{r})$, $p(\mathbf{r})$, $\mathbf{u}(\mathbf{r})$\} defined on a domain $\Omega$, compute local approximations \{$\langle\rho_{h}(\mathbf{r})\rangle$, $\langle p_{h}(\mathbf{r})\rangle$,$\widetilde{\mathbf{u}}_{h}(\mathbf{r})$\} which faithfully represent the behavior of the continuum field on scales above some, user defined, filter length (here denoted $h$) and which truncates scales smaller than $\mathcal{O}(h)$. 

The filtering procedure is chosen so as to derive a filtered form of the compressible Navier-Stokes equations (CNSEs) that are consistent with the explicit LES model. This is defined as the filtering integral transform (FIT) and its application to the CNSEs is discussed in \cite{Cholas2018}.  
\begin{myprop}[FIT for fluids]\label{prop:conv}
	Let $\Omega_{h}(\mathbf{r})$ be a locally compact space within the fluid domain $\Omega$. Then the filtered mass density, momentum density and pressure are given by the FIT; for each $w_{h}\in C^{\infty}_{c}(\Omega_{h})$
	\begin{align}
	\langle\rho_{h}(\mathbf{r})\rangle &=\int_{\Omega_{h}(\mathbf{r})}\rho(\mathbf{r}^{\prime})w_{h}(\mathbf{r}-\mathbf{r}^{\prime})d^{\nu}\mathbf{r}^{\prime}\label{deq:l1}\\
	\langle\rho_{h}(\mathbf{r})\rangle\widetilde{\mathbf{u}}_{h}(\mathbf{r}) &=\int_{\Omega_{h}(\mathbf{r})}\rho(\mathbf{r}^{\prime})\mathbf{u}(\mathbf{r}^{\prime})w_{h}(\mathbf{r}-\mathbf{r}^{\prime})d^{\nu}\mathbf{r}^{\prime}\label{deq:1pp}\\
	\langle p_{h}(\mathbf{r})\rangle &=\int_{\Omega_{h}(\mathbf{r})}p(\mathbf{r}^{\prime})w_{h}(\mathbf{r}-\mathbf{r}^{\prime})d^{\nu}\mathbf{r}^{\prime}\label{deq:pp1}
	\end{align} 
\end{myprop}
The smoothed field \{$\langle\rho_{h}(\mathbf{r})\rangle$, $\langle p_{h}(\mathbf{r})\rangle$, $\widetilde{\mathbf{u}}_{h}(\mathbf{r})$\} represents the interaction of fluid particles located at $\mathbf{r}$, $\mathbf{r}^{\prime}\in\Omega_{h}(\mathbf{r})$. Furthermore, the choice of the velocity smoothing here arises from the physical consideration that the smoothed velocity $\widetilde{\mathbf{u}}_{h}:=\langle\mathbf{P}_{h}\rangle/\langle\rho_{h}\rangle$ where $\mathbf{P}$ is the momentum density.

We start with continuum form of the Navier-Stokes equations (NSE) for a compressible fluid describing the time evolution of the disordered field \{$\rho(\mathbf{r})$, $p(\mathbf{r})$, $\mathbf{u}(\mathbf{r})$\}.
\begin{eqnarray}
\frac{d}{dt}\rho&=&-\rho\nabla\cdot\mathbf{u}\label{eqn:201713b}\\
\kappa_{s}(p)\frac{d }{dt}p&=&-\nabla\cdot\mathbf{u} + \gamma\alpha\nabla \cdot\left(\kappa_{s}(p)\nabla p\right)-\alpha\nabla \cdot\left(\frac{1}{\rho}\nabla \rho\right)\label{eq2017:6b}\\
\rho\frac{d}{dt}\mathbf{u}&=&-\nabla p+\nabla\cdot\underline{\underline{\sigma}}+\rho\mathbf{b}\label{eqn:201723b}\\
\frac{d\mathbf{r}}{dt} &= \mathbf{u}\label{deq:tt1}
\end{eqnarray}
with  adiabatic compressibility $\kappa_{s}$, adiabatic incompressibility modulus $K_{s}=1/\kappa_{s}$,  thermal diffusivity $\alpha$, adiabatic index $\gamma$, fluid pressure $p$, fluid density $\rho$, fluid velocity $\mathbf{u}$, body force $\mathbf{b}$ and viscous stress tensor $\underline{\underline{\sigma}}$.   

If the FIT is applied to (\ref{eqn:201713b}), (\ref{eq2017:6b}) and (\ref{eqn:201723b}) we obtain the following set of filtered equations.
\begin{eqnarray}
\frac{d}{dt}\langle\rho_{h}(\mathbf{r})\rangle&=&-\langle\rho_{h}(\mathbf{r})\rangle\nabla\cdot\widetilde{\mathbf{u}}_{h}(\mathbf{r})\label{eqn:2017cc2}\\
\frac{d }{dt}\langle p_{h}(\mathbf{r})\rangle&=&-\langle K_{S}\nabla\cdot\mathbf{u},w_{h}\rangle + \gamma\alpha \langle K_{S}\nabla\cdot(\kappa_{s}\nabla p),w_{h}\rangle\label{eqn:2017c3}\\
\langle\rho_{h}\rangle\frac{d}{dt}\widetilde{\mathbf{u}}_{h}&=&\langle\nabla\cdot \underline{\underline{\tau}},w_{h}\rangle-\nabla\cdot\langle \underline{\underline{\mathcal{H}}}_{h}\rangle+ \langle\rho_{h}\rangle\widetilde{\mathbf{b}}_{h}\label{eqn:2017cc4}\\
\frac{d\mathbf{r}}{dt} &=& \widetilde{\mathbf{u}}_{h}(\mathbf{r})\label{eqn:2017c5}
\end{eqnarray}
where the material derivative after the filtering becomes
\begin{align}
\frac{d}{dt}&=\frac{\partial}{\partial t} + \widetilde{\mathbf{u}}_{h}\cdot\bm{\nabla}
\end{align}
The sub-particle stress (SPS) tensor arising from the filtering process is given by the following definition.
\begin{mydef}[sub-grid stress tensor, SPS]\label{def:sgs}
	Application of the FIT is applied to the momentum equation introduces momentum transfer due to small scale motion. The SPS  represents the effect of the unresolved small scales on the local approximations. This is defined by the following 
	\begin{align}
	\langle\underline{\underline{\mathcal{H}}}_{h}(\mathbf{r})\rangle&=\int_{\Omega(\mathbf{r})}\rho(\mathbf{r}^{\prime})(\mathbf{u}(\mathbf{r}^{\prime})-\widetilde{\mathbf{u}}_{h}(\mathbf{r}))\otimes(\mathbf{u}(\mathbf{r}^{\prime})-\widetilde{\mathbf{u}}_{h}(\mathbf{r}))w_{h}d\Omega(\mathbf{r}^{\prime})\nonumber\\&=\langle\rho_{h}(\mathbf{r})\rangle\left(\widetilde{(\mathbf{u}\otimes\mathbf{u})}_{h}(\mathbf{r})-\widetilde{\mathbf{u}}_{h}(\mathbf{r})\otimes\widetilde{\mathbf{u}}_{h}(\mathbf{r})\right)\label{deq:2018d11}\quad\text{by the FIT}
	\end{align}
\end{mydef}

The main task now is to de-filter the filtered equations (\ref{eqn:2017cc2}), (\ref{eqn:2017c3}), (\ref{eqn:2017cc4}) and (\ref{eqn:2017c5}). To this end, an inverse filtering procedure is necessary.

For a detailed development of the filtering process the reader is referred to our other work \cite{Cholas2018}.

\section{SPH Consistent with implicit LES}
\paragraph{De-filtering problem:} The de-filtering problem can be formally posed as follows;
Given the filtered equations governing the evolution of the local approximations $\{\langle\rho_{h}\rangle, \langle p_{h}\rangle,\widetilde{\mathbf{u}}_{h}\}$, de-filter these averaged equations to find the integro-differential equations governing the underlying disordered field $\{\rho, p,\mathbf{u}\}$. The goal of this de-filtering process is to recover or restore the mechanical information at small scale that is lost during the filtering process. Consequently, any turbulent phenomena will be implicitly modeled in this approach. Therefore, the method will be referred to as SPH-$i$, where the $i$ means implicit; it signifies the fact that this version of SPH is consistent with implicit LES.
\begin{myprop}[DIT for fluids]\label{prop:deconv}
	Consider a fluid particle located at $\mathbf{r}$ and has a test space $\Omega_{h}(\mathbf{r})$ within the fluid domain $\Omega$. Given the locally averaged mass density, momentum density and pressure on $\Omega_{h}(\mathbf{r})$, we can reconstruct the continuum field by de-filtering the filtered mass density, momentum density and pressure in proposition \ref{prop:conv}. Mathematically, for each $w_{h}\in C^{\infty}_{c}(\Omega_{h})$, there exists a $\varphi_{h}\in C^{\infty}_{c}(\Omega_{h})$ such that
	\begin{align}
	\rho(\mathbf{r}) &=\int_{\Omega_{h}(\mathbf{r})}\langle\rho_{h}(\mathbf{r}^{\prime})\rangle\varphi_{h}(\mathbf{r}-\mathbf{r}^{\prime})d^{\nu}\mathbf{r}^{\prime}\label{deq:3}\\
	\rho(\mathbf{r})\mathbf{u}(\mathbf{r}) &=\int_{\Omega_{h}(\mathbf{r})}\langle\rho_{h}(\mathbf{r}^{\prime})\rangle\widetilde{\mathbf{u}}_{h}(\mathbf{r}^{\prime})\varphi_{h}(\mathbf{r}-\mathbf{r}^{\prime})d^{\nu}\mathbf{r}^{\prime}\label{deq:3e}\\
	p(\mathbf{r}) &=\int_{\Omega_{h}(\mathbf{r})}\langle p_{h}(\mathbf{r}^{\prime})\rangle\varphi_{h}(\mathbf{r}-\mathbf{r}^{\prime})d^{\nu}\mathbf{r}^{\prime}\label{deq:3g}
	\end{align} 
\end{myprop} 
We then call $w_{h}$ as the convolution filter and $\varphi_{h}$ as the deconvolution filter.

\subsection{De-filtering the filtered CNSE}
The DIT of proposition \ref{prop:deconv} is now applied to the filtered equations to reconstruct the original flow field  provided that the local approximation $\{\langle\rho_{h}\rangle, \langle p_{h}\rangle,\widetilde{\mathbf{u}}_{h}\}$ still contains enough mechanical information that a de-convolution filter can recover the original underlying field $\{\rho, p,\mathbf{u}\}$. 

In SPH, the target particle moves with the filtered or smoothed velocity. In the context of the proposed SPH-$i$, the target particle moves with the de-filtered  velocity. Consider a test particle located at position $\mathbf{r}$ having a test space $\Omega_{h}(\mathbf{r})$. Let there be a support material particle located at $\mathbf{r}^{\prime}$ so that $\mathbf{r}^{\prime}\in\Omega_{h}(\mathbf{r})$. We define the velocities on this locally compact space as
\begin{eqnarray}
\text{test or target particle:}\quad\frac{d\mathbf{r}}{dt} = \mathbf{u}(\mathbf{r}),\quad\text{support particle:}\quad\frac{d\mathbf{r}^{\prime}}{dt} = \widetilde{\mathbf{u}}_{h}(\mathbf{r}^{\prime})\label{deq:t2}
\end{eqnarray}

Consider the filtered continuity equation given by (\ref{eqn:2017cc2}). To de-filter it, we first consider the continuum point $\mathbf{r}^{\prime}$ with a test space $\Omega_{h}(\mathbf{r}^{\prime})$ for all $\mathbf{r}^{\prime}\in\Omega_{h}(\mathbf{r})$. Using the locally averaged variables on $\Omega_{h}(\mathbf{r}^{\prime})$, the de-filtered continuity equation is then tested with the deconvolution filter as given below.    
\begin{align}
&\int_{\Omega_{h}(\mathbf{r})}\left\{\frac{d}{dt}\langle\rho_{h}(\mathbf{r}^{\prime})\rangle+\langle\rho_{h}(\mathbf{r}^{\prime})\rangle\nabla^{\prime}\cdot\widetilde{\mathbf{u}}_{h}(\mathbf{r}^{\prime})\right\}\varphi_{h}(\mathbf{r}-\mathbf{r}^{\prime})d^{\nu}\mathbf{r}^{\prime}=0, \nonumber \\ & \quad {} ^\forall w_{h}\in C^{\infty}_{c}(\Omega_{h}), \quad ^\exists\varphi_{h}\in C^{\infty}_{c}(\Omega_{h})\label{deq:6}
\end{align}
This can be rearranged into suitable form yielding
\begin{align}
&\int_{\Omega_{h}(\mathbf{r})}\left\{\frac{d}{dt}\bigg(\langle\rho_{h}(\mathbf{r}^{\prime})\rangle\varphi_{h}\bigg)+\bigg(\langle\rho_{h}(\mathbf{r}^{\prime})\rangle\varphi_{h}\bigg)\nabla^{\prime}\cdot\widetilde{\mathbf{u}}_{h}(\mathbf{r}^{\prime})\right\}d^{\nu}\mathbf{r}^{\prime}, \nonumber \\ &= \int_{\Omega_{h}(\mathbf{r})}\langle\rho_{h}(\mathbf{r}^{\prime})\rangle\frac{d}{dt}\varphi_{h}(\mathbf{r}-\mathbf{r}^{\prime})d^{\nu}\mathbf{r}^{\prime}\label{deq:6x}
\end{align}
In this form, Reynolds' Transport Theorem is applied to the left hand side and the chain rule of differentiation to the right hand side. Accordingly,
\begin{align}
&\frac{d}{dt}\int_{\Omega_{h}(\mathbf{r})}\langle\rho_{h}(\mathbf{r}^{\prime})\rangle\varphi_{h}(\mathbf{r}-\mathbf{r}^{\prime})d^{\nu}\mathbf{r}^{\prime} \nonumber \\ & \quad =\int_{\Omega_{h}(\mathbf{r})}\langle\rho_{h}(\mathbf{r}^{\prime})\rangle\frac{d}{dt}\varphi_{h}(\mathbf{r}-\mathbf{r}^{\prime})d^{\nu}\mathbf{r}^{\prime}\nonumber \\ & \quad {}
=\int_{\Omega_{h}(\mathbf{r})}\langle\rho_{h}(\mathbf{r}^{\prime})\rangle \left\{\frac{d\mathbf{r}}{dt}\cdot\nabla\varphi_{h}+\frac{d\mathbf{r}^{\prime}}{dt}\cdot\nabla^{\prime}\varphi_{h}\right\}d^{\nu}\mathbf{r}^{\prime}\nonumber \\ & \quad {}
=\int_{\Omega_{h}(\mathbf{r})}\langle\rho_{h}(\mathbf{r}^{\prime})\rangle \left\{\mathbf{u}(\mathbf{r})\cdot\nabla\varphi_{h}+\widetilde{\mathbf{u}}_{h}(\mathbf{r}^{\prime})\cdot\nabla^{\prime}\varphi_{h}\right\}d^{\nu}\mathbf{r}^{\prime}\nonumber \\ & \quad {} 
=\int_{\Omega_{h}(\mathbf{r})}\langle\rho_{h}(\mathbf{r}^{\prime})\rangle \bigg(\mathbf{u}(\mathbf{r})-\widetilde{\mathbf{u}}_{h}(\mathbf{r}^{\prime})\bigg)\cdot\nabla\varphi_{h}d^{\nu}\mathbf{r}^{\prime}\nonumber\\
&\frac{d}{dt}\rho(\mathbf{r})=\int_{\Omega_{h}(\mathbf{r})}\langle\rho_{h}(\mathbf{r}^{\prime})\rangle \bigg(\mathbf{u}(\mathbf{r})-\widetilde{\mathbf{u}}_{h}(\mathbf{r}^{\prime})\bigg)\cdot\nabla\varphi_{h}d^{\nu}\mathbf{r}^{\prime}\label{deq:7}
\end{align}
where the anti-symmetry property of the deconvolution gradient $\nabla^{\prime}\varphi=-\nabla\varphi$ has been used to simplify the above.

Further transformation of (\ref{deq:7}) into differential form leads to the canonical point form of the continuum continuity equation.
\begin{proof}
	We begin by unplugging the space derivatives from the integral in (\ref{deq:7}) and using the DIT of proposition \ref{prop:deconv} to get
	\begin{align}
	\frac{d}{dt}\rho(\mathbf{r})&=\mathbf{u}(\mathbf{r})\cdot\nabla\int_{\Omega_{h}(\mathbf{r})}\langle\rho_{h}(\mathbf{r}^{\prime})\rangle \varphi_{h}(\mathbf{r}-\mathbf{r}^{\prime}) d^{\nu}\mathbf{r}^{\prime} \nonumber \\ & \quad {} -\nabla\cdot\int_{\Omega_{h}(\mathbf{r})}\langle\rho_{h}(\mathbf{r}^{\prime})\rangle\widetilde{\mathbf{u}}_{h}(\mathbf{r}^{\prime})\varphi_{h}(\mathbf{r}-\mathbf{r}^{\prime})d^{\nu}\mathbf{r}^{\prime}\nonumber \\ & \quad {}
	=\mathbf{u}(\mathbf{r})\cdot\nabla\rho(\mathbf{r}) \nonumber-\nabla\cdot\big(\rho(\mathbf{r})\mathbf{u}(\mathbf{r})\big)\quad\text{by the DIT}\nonumber \\ &\quad {}=-\rho(\mathbf{r})\nabla\cdot\mathbf{u}(\mathbf{r})\label{deq:9}
	\end{align}
	and we recover the point form of the continuum continuity equation.
\end{proof}

Note that (\ref{deq:7}) and (\ref{deq:9}) both represent the continuum form of the continuity equation.
This equivalence immediately leads to the following corollary.
\begin{corollary}[de-filtered velocity divergence]
	Due to the equivalence  of (\ref{deq:7}) and (\ref{deq:9}), the velocity divergence in a continuum can be calculated as an integral
	\begin{align}
	\rho(\mathbf{r})\nabla\cdot\mathbf{u}(\mathbf{r})&=-\int_{\Omega_{h}(\mathbf{r})}\langle\rho_{h}(\mathbf{r}^{\prime})\rangle \bigg(\mathbf{u}(\mathbf{r})-\widetilde{\mathbf{u}}_{h}(\mathbf{r}^{\prime})\bigg)\cdot\nabla\varphi_{h}d^{\nu}\mathbf{r}^{\prime}\label{deq:4c}
	\end{align}
\end{corollary}

We emphasize that (\ref{deq:4c}) is the most fundamental result of the de-filtering process. It will be used to generate momentum conserving integral representations of the pressure gradient and divergence of the stress tensor.

\subsection{De-filtered momentum equation}
De-filtering the smoothed momentum (\ref{eqn:2017cc4}) is a bit more involving than that of the continuity equation due to the presence of sub-grid stresses. It is prudent to clearly demonstrate how the sub-particle stress (SPS) tensor vanishes after the de-filtering process. Consequently, any sub-scale phenomena will be implicitly modeled- a concept that underpins the development of implicit LES models. With this understanding, the proposed iSPH model does not require any turbulence modeling (the i in iSPH signifies that any turbulent phenomena is implicitly captured).   

First consider a fluid particle at the continuum point $\mathbf{r}^{\prime}$ with a test space $\Omega_{h}(\mathbf{r}^{\prime})$ for all $\mathbf{r}^{\prime}\in\Omega_{h}(\mathbf{r})$. The filtered momentum equation is then tested with the deconvolution kernel
\begin{align}
&\int_{\Omega_{h}(\mathbf{r})}\bigg\{\langle\rho_{h}(\mathbf{r}^{\prime})\rangle\frac{d}{dt}\widetilde{\mathbf{u}}_{h}(\mathbf{r}^{\prime})-\langle\nabla^{\prime}\cdot \underline{\underline{\tau}}(\mathbf{r}^{\prime}),w_{h}\rangle+\nabla^{\prime}\cdot\langle \underline{\underline{\mathcal{H}}}_{h}(\mathbf{r}^{\prime})\rangle+ \nonumber\\ & \qquad \langle\rho_{h}(\mathbf{r}^{\prime})\rangle\widetilde{\mathbf{b}}_{h}(\mathbf{r}^{\prime})\bigg\}\varphi(\mathbf{r}-\mathbf{r}^{\prime})d^{\nu}\mathbf{r}^{\prime} = 0\quad ^\forall w_{h}\in C^{\infty}_{c}(\Omega_{h}),~^\exists\varphi_{h}\in C^{\infty}_{c}(\Omega_{h})
\label{deq:d1}
\end{align}
which can be re-written as
\begin{align}
&\int_{\Omega_{h}(\mathbf{r})}\langle\rho_{h}(\mathbf{r}^{\prime})\rangle\frac{d}{dt}\bigg(\widetilde{\mathbf{u}}_{h}(\mathbf{r}^{\prime})\varphi_{h}\bigg)d^{\nu}\mathbf{r}^{\prime}=\int_{\Omega_{h}(\mathbf{r})}\langle\rho_{h}(\mathbf{r}^{\prime})\rangle\widetilde{\mathbf{u}}_{h}(\mathbf{r}^{\prime})\frac{d\varphi_{h}}{dt}d^{\nu}\mathbf{r}^{\prime}\nonumber\\ &\qquad+\int_{\Omega_{h}(\mathbf{r})}\nabla^{\prime\prime}\cdot \underline{\underline{\tau}}(\mathbf{r}^{\prime\prime})\left(\int_{\Omega_{h}(\mathbf{r}^{\prime})}w_{h}(\mathbf{r}^{\prime\prime}-\mathbf{r}^{\prime})\varphi(\mathbf{r}-\mathbf{r}^{\prime})d^{\nu}\mathbf{r}^{\prime}\right)d^{\nu}\mathbf{r}^{\prime\prime}\nonumber\\ &\qquad-\int_{\Omega_{h}(\mathbf{r})}\nabla^{\prime}\cdot\langle \underline{\underline{\mathcal{H}}}_{h}(\mathbf{r}^{\prime})\rangle\varphi(\mathbf{r}-\mathbf{r}^{\prime})d^{\nu}\mathbf{r}^{\prime}\nonumber\\ & \qquad +\int_{\Omega_{h}(\mathbf{r})}\langle\rho_{h}(\mathbf{r}^{\prime})\rangle\widetilde{\mathbf{b}}_{h}(\mathbf{r}^{\prime})\varphi(\mathbf{r}-\mathbf{r}^{\prime})d^{\nu}\mathbf{r}^{\prime}
\label{deq:d2}
\end{align}
note use of the completeness statement \ref{deq:12} to simplify the second on the right hand side. 
\begin{align}
&\int_{\Omega_{h}(\mathbf{r})}\langle\rho_{h}(\mathbf{r}^{\prime})\rangle\frac{d}{dt}\left(\widetilde{\mathbf{u}}_{h}(\mathbf{r}^{\prime})\varphi_{h}\right)d^{\nu}\mathbf{r}^{\prime}=\int_{\Omega_{h}(\mathbf{r})}\langle\rho_{h}(\mathbf{r}^{\prime})\rangle\widetilde{\mathbf{u}}_{h}(\mathbf{r}^{\prime})\frac{d}{dt}\varphi_{h}d^{\nu}\mathbf{r}^{\prime}+\rho\mathbf{b}\nonumber\\ &\qquad+\int_{\Omega_{h}(\mathbf{r})}\nabla^{\prime\prime}\cdot \underline{\underline{\tau}}(\mathbf{r}^{\prime\prime})\delta(\mathbf{r}-\mathbf{r}^{\prime})d^{\nu}\mathbf{r}^{\prime\prime}-\int_{\Omega_{h}(\mathbf{r})}\nabla^{\prime}\cdot\langle \underline{\underline{\mathcal{H}}}_{h}(\mathbf{r}^{\prime})\rangle\varphi(\mathbf{r}-\mathbf{r}^{\prime})d^{\nu}\mathbf{r}^{\prime}\label{deq:d5}
\end{align}
Once again by applying the Reynolds transport theorem we obtain
\begin{align}
&\frac{d}{dt}\int_{\Omega_{h}(\mathbf{r})}\langle\rho_{h}(\mathbf{r}^{\prime})\rangle\widetilde{\mathbf{u}}_{h}(\mathbf{r}^{\prime})\varphi_{h}d^{\nu}\mathbf{r}^{\prime}=\int_{\Omega_{h}(\mathbf{r})}\langle\rho_{h}(\mathbf{r}^{\prime})\rangle\widetilde{\mathbf{u}}_{h}(\mathbf{r}^{\prime})\frac{d}{dt}\varphi_{h}d^{\nu}\mathbf{r}^{\prime}\nonumber\\ &\qquad+ \nabla\cdot\underline{\underline{\tau}}(\mathbf{r})-\int_{\Omega_{h}(\mathbf{r})}\nabla^{\prime}\cdot\langle \underline{\underline{\mathcal{H}}}_{h}(\mathbf{r}^{\prime})\rangle\varphi(\mathbf{r}-\mathbf{r}^{\prime})d^{\nu}\mathbf{r}^{\prime}+\rho(\mathbf{r})\mathbf{b}(\mathbf{r})\label{deq:d7}
\end{align}

By further applying the DIT to the left hand side and the chain rule of differentiation to the first term on the right hand side the following simplified integro-differential equation is to obtained. 
\begin{align}
&\rho(\mathbf{r})\frac{d\mathbf{u}(\mathbf{r})}{dt}=-\int_{\Omega_{h}(\mathbf{r})}\langle\rho_{h}(\mathbf{r}^{\prime})\rangle(\mathbf{u}(\mathbf{r})-\widetilde{\mathbf{u}}_{h}(\mathbf{r}^{\prime}))\otimes(\mathbf{u}(\mathbf{r})-\widetilde{\mathbf{u}}_{h}(\mathbf{r}^{\prime}))\cdot\nabla\varphi_{h}d^{\nu}\mathbf{r}^{\prime}\nonumber\\ &\qquad-\int_{\Omega_{h}(\mathbf{r})}\nabla^{\prime}\cdot\langle \underline{\underline{\mathcal{H}}}_{h}(\mathbf{r}^{\prime})\rangle\varphi(\mathbf{r}-\mathbf{r}^{\prime})d^{\nu}\mathbf{r}^{\prime}+\nabla\cdot\underline{\underline{\tau}}(\mathbf{r})+\rho(\mathbf{r})\mathbf{b}(\mathbf{r})\label{deq:d8}
\end{align}

The next step is to show that the first and second terms on the right hand side of (\ref{deq:d8}) add to zero by noting the following; By expanding the integrand in the first term and applying the DIT yields
\begin{align}
&\int_{\Omega_{h}(\mathbf{r})}\langle\rho_{h}(\mathbf{r}^{\prime})\rangle(\mathbf{u}(\mathbf{r})-\widetilde{\mathbf{u}}_{h}(\mathbf{r}^{\prime}))\otimes(\mathbf{u}(\mathbf{r})-\widetilde{\mathbf{u}}_{h}(\mathbf{r}^{\prime}))\cdot\nabla\varphi_{h}d^{\nu}\mathbf{r}^{\prime}\nonumber\\ &=\nabla\cdot\left(\int_{\Omega_{h}(\mathbf{r})}\langle\rho_{h}(\mathbf{r}^{\prime})\rangle\widetilde{\mathbf{u}}_{h}(\mathbf{r}^{\prime})\otimes\widetilde{\mathbf{u}}_{h}(\mathbf{r}^{\prime})\varphi(\mathbf{r}-\mathbf{r}^{\prime})d^{\nu}\mathbf{r}^{\prime} -\rho(\mathbf{r})\mathbf{u}(\mathbf{r})\otimes\mathbf{u}(\mathbf{r})\right)\label{deq:d8a}
\end{align}

Furthermore, using Gauss' theorem it easy is to show that the second term transforms to
\begin{align}
&\int_{\Omega_{h}(\mathbf{r})}\nabla^{\prime}\cdot\langle \underline{\underline{\mathcal{H}}}_{h}(\mathbf{r}^{\prime})\rangle\varphi(\mathbf{r}-\mathbf{r}^{\prime})d^{\nu}\mathbf{r}^{\prime}\nonumber\\ &=-\nabla\cdot\left(\int_{\Omega_{h}(\mathbf{r})}\langle\rho_{h}(\mathbf{r}^{\prime})\rangle\widetilde{\mathbf{u}}_{h}(\mathbf{r}^{\prime})\otimes\widetilde{\mathbf{u}}_{h}(\mathbf{r}^{\prime})\varphi(\mathbf{r}-\mathbf{r}^{\prime})d^{\nu}\mathbf{r}^{\prime} -\rho(\mathbf{r})\mathbf{u}(\mathbf{r})\otimes\mathbf{u}(\mathbf{r})\right)\nonumber\\
&\qquad+\int_{\partial\Omega_{\Gamma}(\mathbf{r})}\hat{n}(\mathbf{r}^{\prime})\cdot\langle \underline{\underline{\mathcal{H}}}_{h}(\mathbf{r}^{\prime})\rangle\varphi(\mathbf{r}-\mathbf{r}^{\prime})d^{\nu-1}\Gamma(\mathbf{r}^{\prime})\label{deq:d8b}
\end{align}
where $\partial\Omega_{\Gamma}(\mathbf{r})$ is the surface that bounds the test space $\Omega_{h}(\mathbf{r})$.  Therefore, plugging (\ref{deq:d8a}) and (\ref{deq:d8b}) into (\ref{deq:d8})

\begin{align}
&\rho(\mathbf{r})\frac{d\mathbf{u}(\mathbf{r})}{dt}=\nabla\cdot\underline{\underline{\tau}}(\mathbf{r})+\rho(\mathbf{r})\mathbf{b}(\mathbf{r})+\int_{\partial\Omega_{\Gamma}(\mathbf{r})}\hat{n}(\mathbf{r}^{\prime})\cdot\langle \underline{\underline{\mathcal{H}}}_{h}(\mathbf{r}^{\prime})\rangle\varphi(\mathbf{r}-\mathbf{r}^{\prime})d^{\nu-1}\Gamma(\mathbf{r}^{\prime})\label{deq:d10}
\end{align}

In particular if the fluid domain $\Omega$ is bounded by the surface $\partial\Omega$, then provided that $\partial\Omega_{\Gamma}(\mathbf{r})\cap\partial\Omega=\emptyset$, then the surface integral above is identically zero since $\varphi=0$ on $\partial\Omega_{\Gamma}(\mathbf{r})$ by construct. Therefore, for unbounded domains the reconstruction is exact as (\ref{deq:d10}) is now identical to (\ref{eqn:201723b}).

\subsection{Momentum Conserving DIT for the Stress Tensor}
In the foregoing, we exploit the fact that the SPH-$i$ model has a Lagrangian given by Eckart's Lagrangian \cite{Eckart1960}. Linear momentum conservation is fundamental to the long term stability of numerical algorithms. The SPH-$i$ model is non-conserving, but momentum conserving integral operators can be constructed by sacrificing energy conservation. The energy will only be conserved in an approximate sense. Here we construct momentum conserving integral operators for the stress tensor. For a general  and rigorous approach for dissipative systems, the reader is referred to \cite{Fang2009}.

For a continuum of fluid contained within the domain $\Omega$, starting with Eckart's Lagrangian \cite{Eckart1960}
\begin{align}
L &= \int_{\Omega}\left(\frac{1}{2}\vert\vert\mathbf{u}\vert\vert^{2} - u\right)\rho d\Omega\label{eqn:2017dd4}
\end{align}
Then the total energy of the hydrodynamic system is becomes
\begin{align}
E &= \int_{\Omega}\vert\vert\mathbf{u}\vert\vert^{2}\rho d\Omega-L\\
&= \int_{\Omega}\left(\frac{1}{2}\vert\vert\mathbf{u}\vert\vert^{2} + u\right)\rho d\Omega\label{eqn:2017d4}
\end{align}
where $u$ is the specific internal energy of the system. 

With the help of the Reynolds transport theorem, the rate of change of the total energy is given as
\begin{eqnarray}
\frac{d}{dt}E &=& \int_{\Omega}\left(\rho\frac{d\mathbf{u}}{dt}\cdot\mathbf{u} + \rho\frac{du}{dt}\right)d\Omega\nonumber\\
&=&\int_{\Omega}\left(\nabla p\cdot\mathbf{u} - p\nabla\cdot\mathbf{u}\right)d\Omega\label{eqn:2017d5}
\end{eqnarray}
If we now substitute for $\nabla\cdot\mathbf{u}$ from (\ref{deq:4c}), and with further simplifications we obtain
\begin{align}
\frac{d}{dt}E= \int_{\Omega}\left(\nabla p\cdot\mathbf{u} -\frac{p}{\rho}\int_{\Omega_{h}(\mathbf{r})}\langle\rho_{h}(\mathbf{r}^{\prime})\rangle \bigg(\mathbf{u}(\mathbf{r})-\widetilde{\mathbf{u}}_{h}(\mathbf{r}^{\prime})\bigg)\cdot\nabla\varphi_{h}d^{\nu}\mathbf{r}^{\prime}\right)d\Omega\label{eqn:2017d6}
\end{align}

Using the FIT on unbounded domains, we can simplify (\ref{eqn:2017d6}) as
\begin{align}
&\frac{dE}{dt}= \int_{\Omega}\left(\nabla p -\int_{\Omega_{h}(\mathbf{r})}\left(\frac{p(\mathbf{r})}{\rho(\mathbf{r})}\langle\rho_{h}(\mathbf{r}^{\prime})\rangle +\frac{p(\mathbf{r}^{\prime})}{\rho(\mathbf{r}^{\prime})}\langle\rho_{h}(\mathbf{r})\rangle\right)\nabla\varphi_{h} d^{\nu}\mathbf{r}^{\prime}\right)\cdot\mathbf{u}(\mathbf{r})d\Omega\nonumber\\ 
&+\int_{\Omega}\frac{p(\mathbf{r})}{\rho(\mathbf{r})}\nabla\cdot\int_{\Omega_{h}(\mathbf{r})} \langle\rho_{h}(\mathbf{r}^{\prime})\rangle\hat{\mathbf{u}}_{h}(\mathbf{r}^{\prime})\varphi_{h}d^{\nu}\mathbf{r}^{\prime}d\Omega\label{eqn:2017d10}
\end{align} 
With the assumption that momentum transfer due to turbulent fluctuations is negligible, then the second term in (\ref{eqn:2017d10}) can be neglected. Therefore, energy is conserved under such conditions if,
\begin{align}
G(p\vert\varphi) &=\nabla p= \int_{\Omega_{h}(\mathbf{r})}\left(\frac{p(\mathbf{r})}{\rho(\mathbf{r})}\langle\rho_{h}(\mathbf{r}^{\prime})\rangle +\frac{p(\mathbf{r}^{\prime})}{\rho(\mathbf{r}^{\prime})}\langle\rho_{h}(\mathbf{r})\rangle\right)\nabla\varphi_{h} d^{\nu}\mathbf{r}^{\prime}\label{eqn:2017d11}
\end{align}
as the fluid domain $\Omega$ is arbitrary.

The de-filtering integral transform for the pressure gradient $G(p\vert\varphi)$ is clearly anti-symmetric, thus momentum conserving, and is variationally consistent with the integral transform for the velocity divergence $D(\mathbf{u}\vert\varphi)$ defined below.
\begin{align}
D(\mathbf{u}\vert\varphi)&=\nabla\cdot\mathbf{u}(\mathbf{r})=-\frac{1}{\rho(\mathbf{r})}\int_{\Omega_{h}(\mathbf{r})}\langle\rho_{h}(\mathbf{r}^{\prime})\rangle \bigg(\mathbf{u}(\mathbf{r})-\widetilde{\mathbf{u}}_{h}(\mathbf{r}^{\prime})\bigg)\cdot\nabla\varphi_{h}d^{\nu}\mathbf{r}^{\prime}\label{eqn:2017d12}
\end{align} 
Similar momentum conserving de-filtering integral transforms for the divergence of the deviatoric stress tensor $D(\mu,\underline{\underline{\sigma}}\vert\varphi)$ and the Laplacian of the pressure $L(\kappa^{s},p\vert\varphi)$ can be constructed. With brevity this procedure is omitted but the result is given below.
\begin{align}
D(\mu,\underline{\underline{\sigma}}\vert\varphi) &:=\nabla\cdot\underline{\underline{\sigma}}\nonumber\\
&=\int_{\Omega_{h}(\mathbf{r})}\left(\frac{\underline{\underline{\sigma}}(\mathbf{r})}{\rho(\mathbf{r})}\langle\rho_{h}(\mathbf{r}^{\prime})\rangle +\frac{\underline{\underline{\sigma}}(\mathbf{r}^{\prime})}{\rho(\mathbf{r}^{\prime})}\langle\rho_{h}(\mathbf{r})\rangle\right)\cdot\nabla\varphi_{h} d^{\nu}\mathbf{r}^{\prime}\label{eqn:2017dd11}\\
L(\kappa^{s},p\vert\varphi)&:=\nabla\cdot(\kappa^{s}\nabla p)\nonumber\\
&=\frac{1}{2}\int_{\Omega_{h}(\mathbf{r})}\bigg[\bigg(\langle\kappa^{s}_{h}(\mathbf{r})\rangle+\langle\kappa^{s}_{h}(\mathbf{r}^{\prime})\rangle\bigg)\bigg(p(\mathbf{r})-p(\mathbf{r}^{\prime})\bigg)\nonumber\\
&+\bigg(\kappa^{s}(\mathbf{r})+\kappa^{s}(\mathbf{r}^{\prime})\bigg)\bigg(\langle p_{h}(\mathbf{r})\rangle-\langle p_{h}(\mathbf{r}^{\prime})\rangle\bigg)\bigg]\frac{(\mathbf{r}-\mathbf{r}^{\prime})\cdot\nabla\varphi_{h}}{\vert\vert\mathbf{r}-\mathbf{r}^{\prime}\vert\vert^{2}}d^{\nu}\mathbf{r}^{\prime}\label{eqn:2017d13}
\end{align} 
For a rigorous derivation of (\ref{eqn:2017d13}) refer to appendix \ref{ap:a}.
\section{Unsmoothed particle hydrodynamics model}
The de-filtered SPH, SPH$-i$, model is a complete model resulting from the application of the DIT to the filtered CNSEs. Unlike the SPH which uses the zeroth order deconvolution method, SPH$-i$ is based on the general deconvolution method. The mathematical procedure is shown below; steps [1]$\sim$[3] is the convolution operation on the fields $\{\rho, p, \mathbf{u}\}$ to produce local approximations $\{\langle\rho_{h}\rangle, \langle p_{h}\rangle, \widetilde{\mathbf{u}}_{h}\}$. For completeness, in steps [4]$\sim$[6] a deconvolution operation is dynamically performed on the local approximations to reconstruct the original continuum field $\{\rho, p, \mathbf{u}\}$.

We also use the de-filtering integral operators (\ref{eqn:2017d11}), (\ref{eqn:2017d12})and (\ref{eqn:2017d13})   

\begin{enumerate}[label={[\arabic*]}]
	\item smoothed mass density
	\begin{align}
	\langle\rho_{h}(\mathbf{r})\rangle&=\int_{\Omega_{h}(\mathbf{r})}\rho(\mathbf{r}^{\prime})w_{h}(\mathbf{r}-\mathbf{r}^{\prime})d^{\nu}\mathbf{r}^{\prime}\label{deq:d12}\nonumber\\
	&\stackrel{.}{=} \rho(\mathbf{r})-\int_{\Omega_{h}(\mathbf{r})}\bigg(\rho(\mathbf{r})-\rho(\mathbf{r}^{\prime})\bigg)w_{h}(\mathbf{r}-\mathbf{r}^{\prime})d^{\nu}\mathbf{r}^{\prime}
	\end{align}
	\item smoothed pressure
	\begin{align}
	\langle p_{h}(\mathbf{r})\rangle&=\int_{\Omega_{h}(\mathbf{r})}p(\mathbf{r}^{\prime})w_{h}(\mathbf{r}-\mathbf{r}^{\prime})d\Omega(\mathbf{r}^{\prime})\label{deq:d16}\\
	&\stackrel{.}{=}p(\mathbf{r})-\int_{\Omega_{h}(\mathbf{r})}\bigg(p(\mathbf{r})-p(\mathbf{r}^{\prime})\bigg)w_{h}(\mathbf{r}-\mathbf{r}^{\prime})d^{\nu}\mathbf{r}^{\prime}
	\end{align}
	\item smoothed velocity
	\begin{align}
	\widetilde{\mathbf{u}}_{h}(\mathbf{r})&=\frac{1}{\langle\rho_{h}(\mathbf{r})\rangle}\int_{\Omega_{h}(\mathbf{r})}\rho(\mathbf{r}^{\prime})\mathbf{u}(\mathbf{r}^{\prime})w_{h}(\mathbf{r}-\mathbf{r}^{\prime})d\Omega(\mathbf{r}^{\prime})\label{deq:d13}\\
	&\stackrel{.}{=}\mathbf{u}(\mathbf{r})-\frac{1}{\langle\rho_{h}(\mathbf{r})\rangle}\int_{\Omega_{h}(\mathbf{r})}\rho(\mathbf{r}^{\prime})\bigg(\mathbf{u}(\mathbf{r})-\mathbf{u}(\mathbf{r}^{\prime})\bigg)w_{h}(\mathbf{r}-\mathbf{r}^{\prime})d^{\nu}\mathbf{r}^{\prime}
	\end{align}
	\item de-filtered continuity equation
	\begin{align}
	\frac{d\rho}{dt}&=-\rho D(\mathbf{u}\vert\varphi)\label{deq:d14}
	\end{align}
	\item de-filtered pressure equation
	\begin{align}
	\kappa_{s}(p)\frac{dp}{dt} &=
	-D(\mathbf{u}\vert\varphi)\nonumber \\
	& \qquad{} +\gamma\alpha L(\kappa_{s}(p),p\vert\varphi)-\alpha L\left(\frac{1}{\rho},\rho\vert\varphi\right)
	\end{align}
	\item de-filtered momentum equation
	\begin{align}
	\rho\frac{d\mathbf{u}}{dt}&=-G(p\vert\varphi)+D(\mu,\underline{\underline{\sigma}}\vert\varphi)+G\left(\underline{\underline{\sigma}}\vert\varphi\right)+\rho\mathbf{b}\label{deq:d15}
	\end{align}
	\item moving the particles
	\begin{align}
	\frac{d\mathbf{r}}{dt} &= \mathbf{u}(\mathbf{r})
	\end{align}
\end{enumerate}

To get the discrete forms we just replace integrals by summations. The reader must also see that the differential forms of the above are the original compressible Navier-Stokes equations.

This model is incomplete without a proper construct of the deconvolution filter. We present a procedure for constructing a deconvolution filter given a convolution filter.

\section{Constructing compatible convolution $\&$ deconvolution filters}
\subsection{Completeness of filtering and de-filtering processes: integral form}\label{subsub:convdeconv}
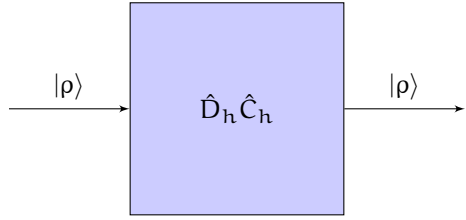
\begin{figure}[H]
	\centering
	\tikzstyle{int}=[draw, fill=blue!20, minimum size=8em]
	\tikzstyle{init} = [pin edge={to-,thin,black}]
	\begin{tikzpicture}[node distance=5.5cm,auto,>=latex']
	\node [int] (a) {$\hat{D}_{h}\hat{C}_{h}$};
	\node (b) [left of=a,node distance=3cm, coordinate] {a};
	%	\node [int, pin={[init]above:$deconvolution~operator$}] (c) [right of=a] {$\hat{D}_{h}$};
	\node [coordinate] (end) [right of=a, node distance=3cm]{};
	\path[->] (b) edge node {$\vert\rho\rangle$} (a);
	\draw[->] (a) edge node {$\vert\rho\rangle$} (end);
	%	\draw[->] (c) edge node {$\vert\rho\rangle$} (end) ;
	\end{tikzpicture}
	
	\caption[Exact resolution of identity using two commuting operators]{Mathematically accurate and consistent completeness property. We exploit the existence of invertible operators to resolve identity.}
	\label{fig:UphJoint}
\end{figure}

A deconvolution operator $\hat{D}_{h}$ exists if the action of the convolution operator $\hat{C}_{h}$ on $\vert\rho\rangle$ i.e. $\hat{C}_{h}\vert\rho\rangle=\vert\bar{\rho}_{h}\rangle$ does not result in irreparable damage so that $\vert\bar{\rho}_{h}\rangle$ still contains enough information that the linear operator $\hat{D}_{h}$ can restore the original input vector $\vert\rho\rangle$ to give back identity i.e. $\hat{D}_{h}\vert\bar{\rho}_{h}\rangle=\vert\rho\rangle$. Figure \ref{fig:UphJoint} denotes completeness without any approximations.
Expressed mathematically in operator space,
\begin{align}
\vert\rho\rangle &=\hat{D}_{h}\hat{C}_{h}\vert\rho\rangle\label{deq:10}
\end{align}
Furthermore, by taking an overlap with the bra vector $\langle\mathbf{r}\vert$ and inserting complete sets of states we obtain a statement of completeness of filtering and de-filtering processes. Then for all $\mathbf{r},\mathbf{r}^{\prime},\mathbf{r}^{\prime\prime}\in\Omega_{h}(\mathbf{r})$ we have
\begin{align}
\langle \mathbf{r}\vert\rho\rangle
&=\langle \mathbf{r}\vert\hat{D}_{h}\hat{1}\hat{C}_{h}\hat{1}\vert\rho\rangle\nonumber\\ 
&=\langle \mathbf{r}\vert\hat{D}_{h}\int_{\Omega_{h}(\mathbf{r})}\vert \mathbf{r}^{\prime}\rangle\langle \mathbf{r}^{\prime}\vert d\Omega(\mathbf{r}^{\prime})\hat{C}_{h}\int_{\Omega_{h}(\mathbf{r}^{\prime})}\vert \mathbf{r}^{\prime\prime}\rangle\langle \mathbf{r}^{\prime\prime}\vert d\Omega(\mathbf{r}^{\prime\prime})\vert\rho\rangle\nonumber\\ 
&= \int_{\Omega_{h}(\mathbf{r})}\int_{\Omega_{h}(\mathbf{r}^{\prime})}\langle \mathbf{r}\vert\hat{D}_{h}\vert \mathbf{r}^{\prime}\rangle\langle \mathbf{r}^{\prime}\vert\hat{C}_{h}\vert \mathbf{r}^{\prime\prime}\rangle\langle \mathbf{r}^{\prime\prime}\vert\rho\rangle d\Omega(\mathbf{r}^{\prime})d\Omega(\mathbf{r}^{\prime\prime})\nonumber\\
\rho(\mathbf{r})&= \int_{\Omega_{h}(\mathbf{r})}\rho (\mathbf{r}^{\prime\prime})\left(\int_{\Omega_{h}(\mathbf{r}^{\prime})}\varphi_{h}(\mathbf{r}-\mathbf{r}^{\prime})w_{h}(\mathbf{r}^{\prime\prime}-\mathbf{r}^{\prime}) d\Omega(\mathbf{r}^{\prime})\right)d\Omega(\mathbf{r}^{\prime\prime})\label{deq:11}
\end{align}
For perfect reconstruction of the underlying continuum field by de-filtering the filtered field, we have the following statement of completeness; for a given convolution filter $w_{h}\in C^{\infty}_{c}(\Omega_{h})$ there exists a deconvolution filter $\varphi_{h}\in C^{\infty}_{c}(\Omega_{h})$ such that
\begin{align}
\int_{\Omega_{h}(\mathbf{r}^{\prime})}\varphi_{h}(\mathbf{r}-\mathbf{r}^{\prime})w_{h}(\mathbf{r}^{\prime\prime}-\mathbf{r}^{\prime}) d\Omega(\mathbf{r}^{\prime}) &= \delta(\mathbf{r}-\mathbf{r}^{\prime\prime})\label{deq:12}
\end{align}
Equation (\ref{deq:12}) represents a fundamental result of the theory which will be used in the construction of explicit, compatible convolution-deconvolution filter pairs.

\begin{figure}[H]
	\centering
	\tikzstyle{int}=[draw, fill=blue!20, minimum size=8em]
	\tikzstyle{init} = [pin edge={to-,thin,black}]
	\begin{tikzpicture}[node distance=5.5cm,auto,>=latex']
	\node [int, pin={[init]above:convolution~operator}] (a) {$\hat{C}_{h}$};
	\node (b) [left of=a,node distance=3cm, coordinate] {a};
	\node [int, pin={[init]above:deconvolution~operator}] (c) [right of=a] {$\hat{D}_{h}$};
	\node [coordinate] (end) [right of=c, node distance=3cm]{};
	\path[->] (b) edge node {$\vert\rho\rangle$} (a);
	\path[->] (a) edge node {$\vert\bar{\rho}_{h}\rangle$} (c);
	\draw[->] (c) edge node {$\vert\rho\rangle$} (end) ;
	\end{tikzpicture}
	
	\caption[Convolution and deconvolution operators in series]{In this series representation perfect reconstruction is realized by filtering and then de-filtering.}
	\label{fig:uphConcept}
\end{figure}
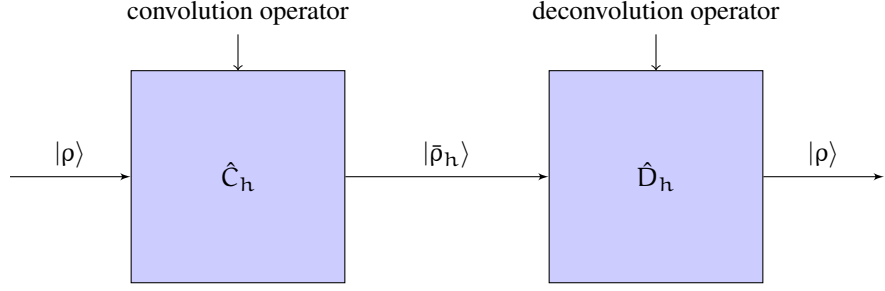

Furthermore, the block diagram \ref{fig:UphJoint} can be cascaded in order to determine the effect of each operator on the input vector. This is depicted in figure \ref{fig:uphConcept}.

Next, we prove that the above FIT and DIT indeed form an integral transform pair through which (perfect) reconstruction of the continuum field is achievable. 

\begin{proof}
	Let $\Omega_{h}(\mathbf{r})$ be a locally compact test space centered at the continuum point $\mathbf{r}$. Consider the continuum point $\mathbf{r}^{\prime}$ with a test space $\Omega_{h}(\mathbf{r}^{\prime})$ for all $\mathbf{r}^{\prime}\in\Omega_{h}(\mathbf{r})$. We first compute the filtered mass density about $\mathbf{r}^{\prime}$ using the FIT; for all $w_{h}\in C^{\infty}_{c}(\Omega_{h})$ we have
	\begin{align}
	\langle\rho_{h}(\mathbf{r}^{\prime})\rangle &=\int_{\Omega_{h}(\mathbf{r}^{\prime})}\rho(\mathbf{r}^{\prime\prime})w_{h}(\mathbf{r}^{\prime}-\mathbf{r}^{\prime\prime})d\Omega(\mathbf{r}^{\prime\prime})\label{deq:4}
	\end{align} 
	Then multiplying (\ref{deq:4}) by the deconvolution filter and integrating over the test space $\Omega_{h}(\mathbf{r})$ yields
	\begin{align}
	&\int_{\Omega_{h}(\mathbf{r})}\langle\rho_{h}(\mathbf{r}^{\prime})\rangle  \varphi_{h}(\mathbf{r}-\mathbf{r}^{\prime})d\Omega(\mathbf{r}^{\prime}) \nonumber \\ & \quad {} =\int_{\Omega_{h}(\mathbf{r})}\int_{\Omega_{h}(\mathbf{r}^{\prime})}\rho(\mathbf{r}^{\prime\prime})w_{h}(\mathbf{r}^{\prime\prime}-\mathbf{r}^{\prime})\varphi_{h}(\mathbf{r}-\mathbf{r}^{\prime})d\Omega(\mathbf{r}^{\prime\prime})d\Omega(\mathbf{r}^{\prime})\nonumber \\ & \quad {}
	=\int_{\Omega_{h}(\mathbf{r})}\rho(\mathbf{r}^{\prime\prime})\left\{\int_{\Omega_{h}(\mathbf{r}^{\prime})}w_{h}(\mathbf{r}^{\prime\prime}-\mathbf{r}^{\prime})\varphi_{h}(\mathbf{r}-\mathbf{r}^{\prime})d\Omega(\mathbf{r}^{\prime})\right\}d\Omega(\mathbf{r}^{\prime\prime})\nonumber \\ & \quad {} 
	=\int_{\Omega_{h}(\mathbf{r})}\rho(\mathbf{r}^{\prime\prime})\delta(\mathbf{r}-\mathbf{r}^{\prime\prime})d\Omega(\mathbf{r}^{\prime\prime})\quad\text{by (\ref{deq:12})}\nonumber \\ & \quad {}
	=\rho(\mathbf{r})\label{deq:5}
	\end{align}
	hence confirming the claim that the FIT of proposition \ref{prop:conv} forms an integral transform pair with the DIT given by proposition \ref{prop:deconv}..
\end{proof}

\begin{figure}[H]
	\centering
	\begin{tikzpicture}
	\shade[shading=radial] (0,5) circle (2);
	\draw[red,fill=red] (0,5) circle (.5ex);
	\draw[green,fill=green] (1,4) circle (.5ex);
	\draw[yellow,fill=yellow] (-1,6) circle (.5ex);
	\draw[blue,fill=blue] (-1.5,4.3) circle (.5ex);
	\draw[brown,fill=brown] (0.8,5.5) circle (.5ex);
	
	\shade[shading=radial] (7,5) circle (2);
	\draw[green] (8,4) circle (2);
	\draw[blue] (5.5,4.3) circle (2);
	\draw[yellow] (6,6) circle (2);
	\draw[brown] (7.8,5.5) circle (2);
	\draw[red,fill=red] (7,5) circle (.5ex);
	\draw[green,fill=green] (8,4) circle (.5ex);
	\draw[yellow,fill=yellow] (6,6) circle (.5ex);
	\draw[blue,fill=blue] (5.5,4.3) circle (.5ex);
	\draw[brown,fill=brown] (7.8,5.5) circle (.5ex);
	\draw[black,fill=black] (5.2,4.3) circle (.5ex);
	\draw[black,fill=black] (4.4,3.5) circle (.5ex);
	\draw[black,fill=black] (5.2,3.1) circle (.5ex);
	\draw[black,fill=black] (4.8,5.3) circle (.5ex);
	\draw[black,fill=black] (5.2,6.6) circle (.5ex);
	\draw[black,fill=black] (7.5,7.1) circle (.5ex);
	\draw[black,fill=black] (9.2,5.3) circle (.5ex);
	\draw[black,fill=black] (9.4,6.5) circle (.5ex);
	\draw[black,fill=black] (9.2,2.6) circle (.5ex);
	\draw[black,fill=black] (7.8,3.1) circle (.5ex);
	
	\draw[black,fill=black] (-1.8,4.3) circle (.5ex);
	\draw[black,fill=black] (-2.6,3.5) circle (.5ex);
	\draw[black,fill=black] (-1.8,3.1) circle (.5ex);
	\draw[black,fill=black] (-2.2,5.3) circle (.5ex);
	\draw[black,fill=black] (-1.8,6.6) circle (.5ex);
	\draw[black,fill=black] (0.5,7.1) circle (.5ex);
	\draw[black,fill=black] (2.2,5.3) circle (.5ex);
	\draw[black,fill=black] (2.4,6.5) circle (.5ex);
	\draw[black,fill=black] (2.2,2.6) circle (.5ex);
	\draw[black,fill=black] (0.8,3.1) circle (.5ex);
	\end{tikzpicture}	
	\caption[Physical interpretation of FIT and DIT]{Discrete interpretation of Completeness. Left hand side: discrete of FIT is a gathering process. Right hand side: discrete DIT is a scattering process. Here $\textcolor{red}{\bullet}$ is the target particle and has support particles $\{\textcolor{blue}{\bullet},\textcolor{yellow}{\bullet},\textcolor{green}{\bullet},\textcolor{brown}{\bullet}\}$. All black particles $\textcolor{black}{\bullet}$ are outside the support of the target particle. The respective domain of influence of each support particle is shown by a circle of corresponding color to that particle.}
	\label{fig:FDConcept}
\end{figure}
%[Physical interpretation of discrete completeness]

We now consider the following sampling problem: Assuming that the density $\rho(\mathbf{r}_{j}):=\rho_{j}$ of each support particle is known, how do we use this information to determine the density $\rho_{i}$ of the $i^{\text{th}}$ target particle?

\paragraph{FIT is a "gather" process} The left hand side of figure \ref{fig:FDConcept} shows the test or target particle $\textcolor{red}{\bullet}$ with support particles 
$\{\textcolor{blue}{\bullet},\textcolor{yellow}{\bullet},\textcolor{green}{\bullet},\textcolor{brown}{\bullet}\}$. By the proof above, we first determine the local density approximation for each support particle $\langle\rho^{h}_{j}\rangle$ for all $\mathbf{r}_{k}\in\Omega_{h}(\mathbf{r}_{j})$ by the FIT;
\begin{align}
\langle\rho^{h}(\mathbf{r}_{j})\rangle &=\sum_{k\in\mathcal{N}(j)}\rho(\mathbf{r}_{k}) w_{h}(\mathbf{r}_{j}-\mathbf{r}_{k})d\Omega(\mathbf{r}_{k})\label{deq:2018pp1}
\end{align}
The support particle $j$ \emph{gathers} contributions from all its nearest neighbors $k\in\mathcal{N}(j)$. This is a gathering process carried out on all support particles within the domain of influence $\Omega_{h}(\mathbf{r}_{j})$ of the $j^{\text{th}}$ support particle with $\mathbf{r}_{j}\in\Omega_{h}(\mathbf{r}_{i})$. 

\paragraph{DIT is a "scatter" process} Finally, to determine the actual density of the $i^{\text{th}}$ target particle $\textcolor{red}{\bullet}$, we use the DIT as shown on the right hand side of figure \ref{fig:FDConcept}.
\begin{align}
\rho(\mathbf{r}_{i}) &=\sum_{j\in\mathcal{N}(i)}\langle\rho^{h}(\mathbf{r}_{j})\rangle \varphi^{h}(\mathbf{r}_{i}-\mathbf{r}_{j})d\Omega(\mathbf{r}_{j})\label{deq:2018pp2}
\end{align}
The target particle $i$ collects contributions from all support particles $j\in\mathcal{N}(i)$ which the space $\Omega_{h}(\mathbf{r}_{j})$ \emph{scatters} onto $\Omega_{h}(\mathbf{r}_{i})\ni\mathbf{r}_{j}$. Therefore, the DIT is a scattering process as shown on the right hand side of figure \ref{fig:FDConcept}.

\begin{remark}
	The choice of volume element is worth investigating. While the approximation $d\Omega(\mathbf{r}_{i}):=m_{i}/\rho(\mathbf{r}_{i})$ has been adopted in this work, it makes the method more complicated due to the implicit nature in which the de-filtered variables must be extracted from the filtered variables. An intuitive way is to use the volume element
	\begin{align}
	d\Omega(\mathbf{r}_{i}) &=\frac{1}{\sqrt{\sum_{j=1} \varphi^{h}(\mathbf{r}_{i}-\mathbf{r}_{j})\sum_{j=1} w^{h}(\mathbf{r}_{i}-\mathbf{r}_{j})}}\label{deq:2018ppp2}
	\end{align}
\end{remark}

\subsection{Constructing deconvolution filters on $\mathbb{R}^{2}$}\label{sect:filterConstruction}

Problem: Given a convolution filter $w_{h}\in C^{\infty}_{c}(\Omega_{h})$ that is used to compute local approximations \{$\langle\rho_{h}(\mathbf{r})\rangle$, $\langle p_{h}(\mathbf{r})\rangle$, $\widetilde{\mathbf{u}}_{h}(\mathbf{r})$\} from the continuum field, construct a compatible deconvolution filter $\varphi_{h}\in C^{\infty}_{c}(\Omega_{h})$ that faithfully reconstructs the underlying continuum field $\{\rho, p, \mathbf{u}\}$ from these local approximations.

The approach to this problem was mainly motivated by pioneering work of Germano \cite{Germano1986}, Konstantopoulos et al. \cite{Konstantopoulos1990}, Mary and Rice \cite{Masry1992} and others on differential filters.

\subsection{Translation Operator}
To address the above problem, we extensively exploit the nice property of the translation (Lagrange shift) operator $\hat{T}:L^{2}(\mathbb{R})\to L^{2}(\mathbb{R})$ which is \begin{inlinelist}
	\item well defined
	\item linear
	\item bounded and
	\item unitary.
\end{inlinelist}
In $\nu$-dimensions we have;
\begin{align}
\rho(\mathbf{r}-h\mathbf{r}^{\ast}) :=\hat{T}(-h\mathbf{r}^{\ast})\rho(\mathbf{r})\equiv e^{-h\mathbf{r}^{\ast}\cdot\nabla}\rho(\mathbf{r})\label{deq:2017f1}
\end{align}
provided that the turbulent flow is smooth enough. 
\begin{proof}
	we temporarily regard $\rho(\mathbf{r}-h\mathbf{r}^{\ast})$ as a function of $s$. Let
	\begin{align}
	\Psi(s)&\stackrel{\textup{def}}{=}\rho(\mathbf{r}-sh\mathbf{r}^{\ast})\equiv \rho(\mathbf{v})\quad\mathbf{v}:=\mathbf{r}-sh\mathbf{r}^{\ast}\nonumber\\
	&=\sum_{n=0}^{\infty}\frac{s^{n}}{n!}\Psi^{(n)}(0)\nonumber
	\end{align}
	We need $\Psi(1)\equiv\rho(\mathbf{r}-h\mathbf{r}^{\ast})$. Now,
	\begin{align}
	\Psi^{\prime}(s)&=\frac{\partial\rho(\mathbf{v})}{\partial v^{\alpha}}\frac{\partial v^{\alpha}}{\partial s}\qquad \alpha=1,2,3.\quad (\text{sum over $\alpha$})\nonumber\\
	&=-\frac{\partial\rho(\mathbf{v})}{\partial v^{\alpha}}hx^{\ast}_{\alpha}\nonumber
	\end{align}
	Hence
	\begin{align}
	\Psi^{(n)}(s)&=(-h\mathbf{r}^{\ast}\cdot\nabla_{\mathbf{v}})^{n}\rho(\mathbf{v})\quad\text{giving}\quad\Psi^{(n)}(0)=(-h\mathbf{r}^{\ast}\cdot\nabla_{\mathbf{r}})^{n}\rho(\mathbf{r})\nonumber
	\end{align}
	Then, for $\Psi(1)$ we obtain
	\begin{align}
	\rho(\mathbf{r}-h\mathbf{r}^{\ast})&= \sum_{n=0}^{\infty}\frac{(-h\mathbf{r}^{\ast}\cdot\nabla_{\mathbf{r}})^{n}}{n!}\rho(\mathbf{r})=e^{-h\mathbf{r}^{\ast}\cdot\nabla}\rho(\mathbf{r})\nonumber
	\end{align}
\end{proof}
Furthermore, the aforementioned properties of the translation operator can be proved as follows;
\begin{enumerate}
	\item $\hat{T}(-h\mathbf{r}^{\ast})$ is linear.
	\begin{proof}
		for $f,g\in L^{2}(\mathbb{R})$ and $\alpha,\beta\in\mathbb{R}$
		\begin{align}
		&\hat{T}(-h\mathbf{r}^{\ast})[\alpha f + \beta g](\mathbf{r})=[\alpha f + \beta g](\mathbf{r}-h\mathbf{r}^{\ast})=\alpha f(\mathbf{r}-h\mathbf{r}^{\ast}) + \beta g(\mathbf{r}-h\mathbf{r}^{\ast})\nonumber\\
		& =\alpha\hat{T}(-h\mathbf{r}^{\ast})f(\mathbf{r}) + \beta\hat{T}(-h\mathbf{r}^{\ast})g(\mathbf{r})\nonumber
		\end{align}
		so $\hat{T}(-h\mathbf{r}^{\ast})$ is a linear operator.
	\end{proof}
	\item $\hat{T}(-h\mathbf{r}^{\ast})$ is well-defined. 
	\begin{proof}
		for $\rho\in L^{2}(\mathbb{R}^{n})$
		\begin{align}
		\int_{\mathbb{R}^{n}}h^{n}\vert\hat{T}(-h\mathbf{r}^{\ast})\rho(\mathbf{r})\vert d^{n}(\mathbf{r}^{\ast})&=\int_{\mathbb{R}^{n}}h^{n}\vert\rho(\mathbf{r-h\mathbf{r}^{\ast}})\vert^{2} d^{n}(\mathbf{r}^{\ast})\nonumber\\
		&=\int_{\mathbb{R}^{n}}\vert\rho(\mathbf{r}^{\prime})\vert^{2} d^{n}(\mathbf{r}^{\prime})<\infty\nonumber
		\end{align}
		therefore $\hat{T}(-h\mathbf{r}^{\ast})\rho(\mathbf{r})\in L^{2}(\mathbb{R})$, so $\hat{T}(-h\mathbf{r}^{\ast})$ is well-defined.
	\end{proof}
	\item $\hat{T}(-h\mathbf{r}^{\ast})$ is bounded.
	\begin{proof}
		for $\rho\in L^{2}(\mathbb{R}^{n})$
		\begin{align}
		\vert\vert\hat{T}(-h\mathbf{r}^{\ast})\rho(\mathbf{r})\vert\vert^{2}&=\int_{\mathbb{R}^{n}}h^{n}\vert\hat{T}(-h\mathbf{r}^{\ast})\rho(\mathbf{r})\vert^{2} d^{n}(\mathbf{r}^{\ast})\nonumber\\
		&=\int_{\mathbb{R}^{n}}\vert\rho(\mathbf{r}^{\prime})\vert^{2} d^{n}(\mathbf{r}^{\prime}) \quad\text{by (2) above.}\nonumber\\
		&=\vert\vert\rho(\mathbf{r})\vert\vert^{2}\nonumber\\
		\therefore\vert\vert\hat{T}(-h\mathbf{r}^{\ast})\rho(\mathbf{r})\vert\vert&=\vert\vert\rho(\mathbf{r})\vert\vert\nonumber
		\end{align}
		meaning that $\hat{T}(-h\mathbf{r}^{\ast})$ is bounded (an isometry).
	\end{proof}
\end{enumerate}

\subsection{Convolution Operator}
Consider the equation (\ref{deq:l1}) in the FIT of proposition \ref{prop:conv}.
For FIT with compactly supported filters defined on $\Omega_{h}(\mathbf{r}):=\{\mathbf{r},\mathbf{r}^{\prime}\in\mathbb{R}^{3}\vert~\vert\vert\mathbf{r}-\mathbf{r}^{\prime}\vert\vert\leq h,~w_{h}\geq 0\}$, centered around $\mathbf{r}=(x,y,z)^{\text{T}}$.  The following definition can be adopted for the 3-dimensional case
\begin{align}
&\langle\rho_{h}(x,y,z)\rangle\nonumber\\ &=\int_{-\infty}^{\infty}\int_{-\infty}^{\infty}\int_{-\infty}^{\infty}\rho(x^{\prime},y^{\prime},z^{\prime})w_{h}\left(\vert\vert\left( x-x^{\prime},y-y^{\prime},z-z^{\prime}\right)\vert\vert\right)dx^{\prime}dy^{\prime}dz^{\prime}\nonumber\\
&=\int_{-\infty}^{\infty}\int_{-\infty}^{\infty}\int_{-\infty}^{\infty}\rho(x^{\prime},y^{\prime},z^{\prime})h^{-3}w\left(\bigg\vert\bigg\vert\left(\frac{x-x^{\prime}}{h},\frac{y-y^{\prime}}{h},\frac{z-z^{\prime}}{h}\right)\bigg\vert\bigg\vert\right)dx^{\prime}dy^{\prime}dz^{\prime}\nonumber\\
&=\int_{-\infty}^{\infty}\int_{-\infty}^{\infty}\int_{-\infty}^{\infty}\rho(x-hx^{\ast},y-hy^{\ast},z-z^{\ast})w\left(\vert\vert\left( x^{\ast},y^{\ast},z^{\ast}\right)\vert\vert\right)dx^{\ast}dy^{\ast}dz^{\ast}\label{deq:2017f2}
\end{align}
by invoking a change of variables $hx^{\ast}:=x-x^{\prime}$, $hy^{\ast}:=y-y^{\prime}$ and $hz^{\ast}:=z-z^{\prime}$. In compact form we then have
\begin{align}
\langle\rho_{h}(\mathbf{r})\rangle&=\int_{\mathbb{R}^{\nu}}w\left(\vert\vert \mathbf{r}^{\ast}\vert\vert\right)\rho(\mathbf{r}-h\mathbf{r}^{\ast})d^{\nu}\mathbf{r}^{\ast}\label{deq:2017f3}
\end{align}
showing that convolution is commutative. Finally, combining (\ref{deq:2017f1}) and (\ref{deq:2017f3}) the FIT now transforms into differential form as
\begin{align}
\langle\rho_{h}(\mathbf{r})\rangle&=\left(\int_{\mathbb{R}^{\nu}}w\left(\vert\vert \mathbf{r}^{\ast}\vert\vert\right)\hat{T}(-h\mathbf{r}^{\ast})d^{\nu}\mathbf{r}^{\ast}\right)\rho(\mathbf{r})\equiv C_{w}(\hat{T})\rho(\mathbf{r})\label{deq:2017f4}
\end{align}
where we identify $C_{w}(\hat{T})$ as the convolution operator, which is continuous (and hence bounded) and is compact. 
\begin{align}
C_{w}(\hat{T})&\stackrel{\textup{def}}{=}\int_{\mathbb{R}^{\nu}}w\left(\vert\vert \mathbf{r}^{\ast}\vert\vert\right)\hat{T}(-h\mathbf{r}^{\ast})d^{\nu}\mathbf{r}^{\ast}\label{deq:2017ff4}
\end{align}

We shall explicitly compute $C_{w}(\hat{T})$ for compactly supported filters on $\Omega_{h}(\mathbf{r})\in\mathbb{R}^{2}$ as commonly used in SPH. In particular, since the convolution filter is radially or circularly symmetric, using polar coordinates $\mathbf{r}^{\ast}:=(\vert\vert\mathbf{r}^{\ast}\vert\vert\cos\theta,\vert\vert\mathbf{r}^{\ast}\vert\vert\sin\theta)^{\text{T}}$ with $\vert\vert\mathbf{r}^{\ast}\vert\vert\in[0,2]$ and $\theta\in[0,2\pi]$ it is easy to show that the convolution operator becomes 
\begin{align}
C_{w}(\hat{T})&=\Theta_{0}J_{0} + \Theta_{1}J_{1}\frac{h^{2}}{2!}\Delta +\Theta_{2} J_{2}\frac{h^{4}}{4!}\Delta^{2} +\Theta_{3} J_{3}\frac{h^{6}}{6!}\Delta^{3}+...\nonumber\\
&=\sum_{k=0}^{\infty}\Theta_{k}J_{k}\frac{h^{2k}}{(2k)!}\nabla^{2k}\label{deq:2017f5}
\end{align}
Where $\Delta=\nabla^{2}$ is the laplacian operator and the moments of the convolution filter $J_{k}$ and the angular part $\Theta_{k}$ are given by
\begin{align}
\Theta_{k}=\frac{2\pi\Gamma(k+\frac{1}{2})}{\Gamma(k+1)\Gamma(\frac{1}{2})},\quad J_{k} = \int_{0}^{2}\vert\vert\mathbf{r}^{\ast}\vert\vert^{2k+1}w(\vert\vert\mathbf{r}^{\ast}\vert\vert)d\vert\vert\mathbf{r}^{\ast}\vert\vert,\quad k=0,1,2,3,...\label{deq:2017f6}
\end{align}
It is important to note that (\ref{deq:2017f4}) is well posed if and only if $\forall k$, $\vert J_{k}\vert<\infty$ meaning that the convolution filter $w_{h}$ must be rapidly decaying in space.

\subsection{Deconvolution Operator}
For the convolution operator given by (\ref{deq:2017f5}), its associated  deconvolution operator is determined from the completeness (\ref{deq:12}) by solving a set of inhomogeneous equations appropriate to the algebra of power series. First, we present the following version of (\ref{deq:12}) suitable for series algebra
\begin{corollary}[Operator form of completeness statement]
	Let $w_{h}\in C_{c}^{\infty}(\Omega_{h})$ be a convolution filter with associated convolution operator $C_{w}(\hat{T})$. Similarly, let $\varphi_{h}\in C_{c}^{\infty}(\Omega_{h})$ be the deconvolution filter with associated  deconvolution operator $D_{w}(\hat{T})$. Then the completeness statement (\ref{deq:12}) can be expressed in operator form as
	\begin{align}
	C_{w}(\hat{T})D_{\varphi}(\hat{T})= 1\Longleftrightarrow D_{\varphi}(\hat{T})C_{w}(\hat{T}) = 1
	\end{align}
	
\end{corollary}
%This claim is proved in section \ref{operator:properties} below.

Using the method of Cauchy products, for a 2D filter, it is easy to show that the deconvolution filter is given by

\begin{align}
D_{\varphi}(\hat{T}):=C^{-1}_{w}(\hat{T})&=\sum_{k=0}^{\infty}M_{k}\frac{h^{2k}}{(2k)!}\nabla^{2k}\label{deq:2017k1}
\end{align}
where the coefficients $\{M_{k}\vert~k=0,1,2,3,\dots\}$ are given by the following infinite dimensional determinant 
\begin{align}
M_{k}&=(-1)^{k}
\begin{vmatrix}
\vdots & \vdots & \vdots & \vdots & \vdots & \vdots & \vdots \\
\dots & 66\Theta_{1}J_{1} & 1 & 0 & 0 & 0 & 0 \\ 
\dots & 990\Theta_{2}J_{2} & 45\Theta_{1}J_{1} & 1 & 0 & 0 & 0 \\ 
\dots & 616\Theta_{3}J_{3} & 210\Theta_{2}J_{2} & 28\Theta_{1}J_{1} & 1 & 0 & 0 \\ 
\dots & 990\Theta_{4}J_{4} & 210\Theta_{3}J_{3} & 70\Theta_{2}J_{2} & 15\Theta_{1}J_{1} & 1 & 0\\ \dots & 66\Theta_{5}J_{5} & 45\Theta_{4}J_{4} & 28\Theta_{3}J_{3} & 15\Theta_{2}J_{2} & 6\Theta_{1}J_{1} & 1\\
\dots & \Theta_{6}J_{6} & \Theta_{5}J_{5} & \Theta_{4}J_{4} & \Theta_{3}J_{3} & \Theta_{2}J_{2} & \Theta_{1}J_{1} 
\end{vmatrix}\label{deq:2018a1}
\end{align} 
and the $k^{\text{th}}$ coefficient can be extracted from the above formula as minor determinants starting from the bottom right element, for example

\begin{align}
M_{1}&=(-1)^{1}
\begin{vmatrix}
\Theta_{1}J_{1} 
\end{vmatrix}
,\quad
M_{2}=(-1)^{2}
\begin{vmatrix}
6\Theta_{1}J_{1} & 1\\
\Theta_{2}J_{2} & \Theta_{1}J_{1} 
\end{vmatrix}
,\quad\nonumber\\
M_{3}&=(-1)^{3}
\begin{vmatrix}
15\Theta_{1}J_{1}&1 & 0\\
15\Theta_{2}J_{2}&6\Theta_{1}J_{1} & 1\\
\Theta_{3}J_{3}&\Theta_{2}J_{2} & \Theta_{1}J_{1} 
\end{vmatrix}
,\dots
\end{align} 

However, we can directly obtain the deconvolution operator from the DIT of proposition \ref{prop:deconv},  Similar to the procedure used for constructing the convolution operator above. For the DIT with compactly supported filters defined on $\Omega_{h}(\mathbf{r})\in\mathbb{R}^{3}$, a compact space centered around $\mathbf{r}=(x,y,z)^{\text{T}}$. Then given the DIT
\begin{align}
\rho(\mathbf{r}) &=\int_{\mathbb{R}^{\nu}}\langle\rho_{h}(\mathbf{r}^{\prime})\rangle\varphi_{h}(\mathbf{r}-\mathbf{r}^{\prime})d^{\nu}\mathbf{r}^{\prime}\nonumber\\
&=\left(\int_{\mathbb{R}^{\nu}}\varphi(\vert\vert\mathbf{r}^{\ast}\vert\vert)\hat{T}(-h\mathbf{r}^{\ast})d^{\nu}\mathbf{r}^{\ast}\right)\langle\rho_{h}(\mathbf{r})\rangle,\quad\text{set $h\mathbf{r}^{\ast}:=\mathbf{r}-\mathbf{r}^{\prime}$}\nonumber\\
\rho(\mathbf{r}) &=D_{\varphi}(\hat{T})\langle\rho_{h}(\mathbf{r})\rangle\label{deq:2017h0}
\end{align}
where the deconvolution operator $D_{\varphi}(\hat{T})$ is now given by the following series.
\begin{align}
D_{\varphi}(\hat{T})
&=\sum_{k=0}^{\infty}\Theta_{k}L_{k}\frac{h^{2k}}{(2k)!}\nabla^{2k}\label{deq:2017h1}
\end{align}
The deconvolution filter moments $L_{k}$ and the angular part $\Theta_{k}$ are given by
\begin{align}
\Theta_{k}=\frac{2\pi\Gamma(k+\frac{1}{2})}{\Gamma(k+1)\Gamma(\frac{1}{2})},\quad L_{k} = \int_{0}^{2}\vert\vert\mathbf{r}^{\ast}\vert\vert^{2k+1}\varphi(\vert\vert\mathbf{r}^{\ast}\vert\vert)d\vert\vert\mathbf{r}^{\ast}\vert\vert,\quad k=0,1,2,3,...\label{deq:2017h2}
\end{align}
Again, we note that (\ref{deq:2017h0}) is well posed if and only if $\forall k$, $\vert L_{k}\vert<\infty$ meaning that the deconvolution filter $\varphi_{h}$ must also be rapidly decaying in space. 
Since the $w_{h}$ and $\varphi_{h}$ are inverse filters, it follows that (\ref{deq:2017k1}) and(\ref{deq:2017h1}) are equivalent, a fundamental result of this discussion and presented in the following proposition.
\begin{myprop}
	Let the moments of the convolution filter $w_{h}$ be $J_{k}$ and the convolution operator given as $C_{w}(\hat{T})$. Then the moment $L_{k}$ of the associated deconvolution filter $\varphi_{h}$ can be analytically determined without prior knowledge of $\varphi_{h}$. Mathematically,
	\begin{align}
	L_{k} = \frac{M_{k}}{\Theta_{k}}\label{deq:2017k8}
	\end{align}
\end{myprop}

\subsection{convolution \& deconvolution operators for the Gaussian filter}
Consider a special case of the Gaussian filter given by
\begin{align}
w_{h} :=\alpha h^{-\nu}e^{-\vert\vert\mathbf{r}-\mathbf{r}^{\prime}\vert\vert^{2}/h^{2}}\label{deq:2017g1}
\end{align}
The moments of this filter in 2D are; $J_{k} :=\Gamma(k+1)/(2\pi)$, with the angular elements $\Theta_{k}$ given by (\ref{deq:2017f6}). Then the associated convolution and deconvolution operator coefficients are given by
\begin{align}
\Theta_{k}J_{k} =\frac{\Gamma(k+\frac{1}{2})}{\Gamma(\frac{1}{2})}, \quad M_{k} =(-1)^{k}\frac{\Gamma(k+\frac{1}{2})}{\Gamma(\frac{1}{2})}
\end{align}
\begin{align}
C_{w}(\hat{T})&= \sum_{k=0}^{\infty}\frac{\Gamma(k+\frac{1}{2})}{\Gamma(\frac{1}{2})}\frac{h^{2k}}{(2k)!}\nabla^{2k}=\sum_{k=0}^{\infty}\frac{h^{2k}}{4^{k}k!}\nabla^{2k}=e^{\frac{1}{4}h^{2}\nabla^{2}}\\
D_{\varphi}(\hat{T})&= \sum_{k=0}^{\infty}(-1)^{k}\frac{\Gamma(k+\frac{1}{2})}{\Gamma(\frac{1}{2})}\frac{h^{2k}}{(2k)!}\nabla^{2k}=\sum_{k=0}^{\infty}(-1)^{k}\frac{h^{2k}}{4^{k}k!}\nabla^{2k}=e^{-\frac{1}{4}h^{2}\nabla^{2}}\label{deq:2017f7}
\end{align}
as expected. This actually is a very important proof showing that the determinant (\ref{deq:2018a1}) is correct. 

\subsection{matrix coefficients of convolution/deconvolution operators}
We define the convolution filter $w_{h}$ as the "matrix element" of the convolution operator $C_{w}(\hat{T})$.
\begin{eqnarray}
w_{h}(\mathbf{r}-\mathbf{r}^{\prime})&\triangleq&\langle\mathbf{r}^{\prime}\vert C_{w}(\hat{T})\vert\mathbf{r}\rangle\nonumber\\
&=&\langle\mathbf{r}^{\prime}\vert\left(\hat{1} + \sum_{k=1}^{\infty}\Theta_{k}J_{k}\frac{h^{2k}}{(2k)!}\nabla^{2k}\right)\vert\mathbf{r}\rangle\nonumber\\
&=&\langle\mathbf{r}^{\prime}\vert\hat{1}\vert\mathbf{r}\rangle + \sum_{k=1}^{\infty}\Theta_{k}J_{k}\frac{h^{2k}}{(2k)!}\langle\mathbf{r}^{\prime}\nabla^{2k}\vert\mathbf{r}\rangle\nonumber\\
\therefore w_{h}(\mathbf{r}-\mathbf{r}^{\prime})&=&\delta(\mathbf{r}-\mathbf{r}^{\prime})+\sum_{k=1}^{\infty}\Theta_{k}J_{k}\frac{h^{2k}}{(2k)!}\nabla^{2k}\delta(\mathbf{r}-\mathbf{r}^{\prime})\label{deq:6f} 
\end{eqnarray}
Similarly, the deconvolution filter is defined as the "matrix element" of the deconvolution operator with respect to the continuous position basis. Following the same procedure above, the deconvolution filter takes the form
\begin{eqnarray}
\varphi_{h}(\mathbf{r}-\mathbf{r}^{\prime})&\triangleq&\langle\mathbf{r}^{\prime}\vert D_{\varphi}(\hat{T})\vert\mathbf{r}\rangle\nonumber\\
\therefore \varphi_{h}(\mathbf{r}-\mathbf{r}^{\prime})&=&\delta(\mathbf{r}-\mathbf{r}^{\prime})+\sum_{k=1}^{\infty}M_{k}\frac{h^{2k}}{(2k)!}\nabla^{2k}\delta(\mathbf{r}-\mathbf{r}^{\prime})\label{deq:7f} 
\end{eqnarray}
For both operators, in the continuum limit we have the flowing important property
\begin{eqnarray}
\lim\limits_{h\to 0}w_{h}(\mathbf{r}-\mathbf{r}^{\prime})&\equiv&\delta(\mathbf{r}-\mathbf{r}^{\prime}),\quad \lim\limits_{h\to 0}\varphi_{h}(\mathbf{r}-\mathbf{r}^{\prime})\equiv\delta(\mathbf{r}-\mathbf{r}^{\prime})\label{deq:2017e6} 
\end{eqnarray}

In fact it will turn out that the deconvolution filter is shaper and taller than the convolution filter. This implies that the deconvolution filter approximates Dirac's delta function much more accurately than the associated convolution filter, and this has profound consequences on accuracy.
\subsection{Properties of convolution and deconvolution operators}\label{operator:properties}
We study the properties of convolution and deconvolution operators by investigating their action on functionals or generalized functions.
\begin{enumerate}[label={[\arabic*]}]
	\item The action of convolution and deconvolution operators on the Dirac's delta function is to produce the convolution and deconvolution filters respectively. Mathematically,
	\begin{align}
	C_{w}(\hat{T})\delta(\mathbf{r}-\mathbf{r}^{\prime})=w_{h}(\mathbf{r}-\mathbf{r}^{\prime}), \quad D_{\varphi}(\hat{T})\delta(\mathbf{r}-\mathbf{r}^{\prime})=\varphi_{h}(\mathbf{r}-\mathbf{r}^{\prime})\label{deq:2017k2}
	\end{align}
	\begin{proof}
		\begin{align}
		w_{h}(\mathbf{r}-\mathbf{r}^{\prime\prime})&=\int_{\mathbb{R}^{\nu}}w_{h}(\mathbf{r}-\mathbf{r}^{\prime})\delta(\mathbf{r}^{\prime}-\mathbf{r}^{\prime\prime})d^{\nu}\mathbf{r}^{\prime}\quad\text{completeness}\nonumber\\
		&=\int_{\mathbb{R}^{\nu}}w(\mathbf{r}^{\ast})\delta(\mathbf{r}-h\mathbf{r}^{\ast}-\mathbf{r}^{\prime\prime})d^{\nu}\mathbf{r}^{\ast},~\text{change of variables}\nonumber\\
		&=\left(\int_{\mathbb{R}^{\nu}}w(\mathbf{r}^{\ast})\hat{T}(-h\mathbf{r}^{\ast})d^{\nu}\mathbf{r}^{\ast}\right)\delta(\mathbf{r}-\mathbf{r}^{\prime\prime})\nonumber\\
		\therefore w_{h}(\mathbf{r}-\mathbf{r}^{\prime\prime})	&=C_{w}(\hat{T})\delta(\mathbf{r}-\mathbf{r}^{\prime\prime})\nonumber
		\end{align}
		A similar proof for the deconvolution operator follows, hence completing the proof. 
	\end{proof}
	\item The convolution filter is the Green's function of the deconvolution operator, whereas the deconvolution filter is the Green's function of the convolution operator i.e.
	\begin{align}
	C_{w}(\hat{T})\varphi_{h}(\mathbf{r}-\mathbf{r}^{\prime\prime})=\delta(\mathbf{r}-\mathbf{r}^{\prime\prime}), \quad D_{\varphi}(\hat{T}) w(\mathbf{r}-\mathbf{r}^{\prime\prime})=\delta(\mathbf{r}-\mathbf{r}^{\prime\prime})\label{deq:2017k3}
	\end{align}
	Either of these relations is equivalent to the completeness statement (\ref{deq:12}). A proof of this can be directly obtained from (\ref{deq:2017k2}) or directly from (\ref{deq:12}). Following the latter approach
	\begin{proof}
		\begin{align}
		\delta(\mathbf{r}-\mathbf{r}^{\prime\prime})&=\int_{\mathbb{R}^{3}}\varphi_{h}(\mathbf{r}-\mathbf{r}^{\prime})w_{h}(\mathbf{r}^{\prime}-\mathbf{r}^{\prime\prime})d^{3}\mathbf{r}^{\prime}\quad\text{completeness}\nonumber\\
		&=\int_{\mathbb{R}^{3}}\varphi(\mathbf{r}^{\ast})w_{h}(\mathbf{r}-h\mathbf{r}^{\ast}-\mathbf{r}^{\prime\prime})d^{3}\mathbf{r}^{\ast},~\text{change of variables}\nonumber\\
		&=\left(\int_{\mathbb{R}^{3}}\varphi(\mathbf{r}^{\ast})\hat{T}(-h\mathbf{r}^{\ast})d^{3}\mathbf{r}^{\ast}\right)w_{h}(\mathbf{r}-\mathbf{r}^{\prime\prime})\nonumber\\
		\therefore \delta(\mathbf{r}-\mathbf{r}^{\prime\prime})	&=D_{\varphi}(\hat{T}) w_{h}(\mathbf{r}-\mathbf{r}^{\prime\prime})\nonumber
		\end{align}
		For a direct proof from (\ref{deq:2017k2}); we multiply by $D_{\varphi}(\hat{T})$ to the first equation in (\ref{deq:2017k2}) to get $D_{\varphi}(\hat{T})C_{w}(\hat{T})\delta(\mathbf{r}-\mathbf{r}^{\prime})=D_{\varphi}(\hat{T}) w_{h}(\mathbf{r}-\mathbf{r}^{\prime})$ yielding the required result since $D_{\varphi}(\hat{T})C_{w}(\hat{T})=1$.
		
		A similar proof for the deconvolution operator follows, hence completing the proof. 
	\end{proof}
	\item The completeness statement (\ref{deq:12}) can be expressed in operator form as
	\begin{align}
	C_{w}(\hat{T})D_{\varphi}(\hat{T}) =1, \quad D_{\varphi}(\hat{T})C_{w}(\hat{T}) =1\label{deq:2017k10}
	\end{align}
	\begin{proof}
		\begin{align}
		C_{w}(\hat{T})\varphi_{h}(\mathbf{r}-\mathbf{r}^{\prime\prime})&=\delta(\mathbf{r}-\mathbf{r}^{\prime\prime})\quad\text{by (\ref{deq:2017k3}) above}\nonumber\\
		C_{w}(\hat{T})D_{\varphi}(\hat{T})\delta(\mathbf{r}-\mathbf{r}^{\prime\prime})&=\delta(\mathbf{r}-\mathbf{r}^{\prime\prime})\quad\text{by (\ref{deq:2017k2}) above}\nonumber\\
		\therefore C_{w}(\hat{T})D_{\varphi}(\hat{T}) &=1\nonumber
		\end{align}
	\end{proof}
	\item The action of the square convolution operator on the deconvolution filter yields the deconvolution filter. Similarly, the action of the square convolution operator on the deconvolution filter yields the convolution filter.
	\begin{align}
	C_{w}(\hat{T})^{2}\varphi_{h}(\mathbf{r}-\mathbf{r}^{\prime\prime})=w_{h}(\mathbf{r}-\mathbf{r}^{\prime\prime}), \quad D_{\varphi}(\hat{T})^{2} w(\mathbf{r}-\mathbf{r}^{\prime\prime})=\varphi_{h}(\mathbf{r}-\mathbf{r}^{\prime\prime})\label{deq:2017k4}
	\end{align}
	\begin{proof}
		\begin{align}
		C_{w}(\hat{T})^{2}\varphi_{h}(\mathbf{r}-\mathbf{r}^{\prime\prime})&=C_{w}(\hat{T})C_{w}(\hat{T})\varphi_{h}(\mathbf{r}-\mathbf{r}^{\prime\prime})\nonumber\\
		&=C_{w}(\hat{T}) \delta(\mathbf{r}-\mathbf{r}^{\prime\prime})\quad\text{by (\ref{deq:2017k3})}\nonumber\\
		&=w_{h}(\mathbf{r}-\mathbf{r}^{\prime\prime})\quad\text{by (\ref{deq:2017k2})}\nonumber
		\end{align}
		A similar proof can be done for the square deconvolution operator.
	\end{proof}
\end{enumerate}

\subsection{Cauchy product of deconvolution operator with itself}
The Cauchy product is the discrete convolution of two infinite series. Since the square deconvolution operator (\ref{deq:2017k4}) can be interpreted as the discrete convolution of the deconvolution operator with itself, we have,
\begin{align}
D_{\varphi}(\hat{T})^{2}&=D_{\varphi}(\hat{T})\cdot D_{\varphi}(\hat{T})\nonumber\\
&=\left(\sum_{l=0}^{\infty}M_{l}\frac{h^{2l}}{(2l)!}\nabla^{2l}\right)\cdot\left(\sum_{m=0}^{\infty}M_{m}\frac{h^{2m}}{(2m)!}\nabla^{2m}\right)\nonumber\\
\therefore D_{\varphi}(\hat{T})^{2}&=\sum_{k=0}^{\infty}\widetilde{M}_{k}\frac{h^{2k}}{(2k)!}\nabla^{2k}\quad\text{where}\quad \widetilde{M}_{k}=\sum_{j=0}^{k}\binom{2k}{2j}M_{j}M_{k-j}\label{deq:2017k5}
\end{align}
Given a convolution filter $w_{h}$, it then follows from (\ref{deq:2017k4}) and (\ref{deq:2017k5}) that the deconvolution filter is given by the following formula.
\begin{align}
\varphi_{h}(\mathbf{r}-\mathbf{r}^{\prime\prime})&=\sum_{k=0}^{\infty}\widetilde{M}_{k}\frac{h^{2k}}{(2k)!}\nabla^{2k}w_{h}(\mathbf{r}-\mathbf{r}^{\prime\prime})\label{deq:2017k6}
\end{align}
which is an exact deconvolution filter uniquely defined for each specified convolution filter with finite moments. Furthermore, the above series solution is truncated due to the fundamental limitation that computers can only handle finite collections of data. The non-unique approximate deconvolution filter, ADF then becomes
\begin{align}
\varphi_{h,n}(\mathbf{r}-\mathbf{r}^{\prime\prime})&=\sum_{k=0}^{n}\widetilde{M}_{k}\frac{h^{2k}}{(2k)!}\nabla^{2k}w_{h}(\mathbf{r}-\mathbf{r}^{\prime\prime})\equiv\sum_{k=0}^{n}f_{k}(h)\psi_{h,2k}(\mathbf{r}-\mathbf{r}^{\prime\prime})\label{deq:2017k7}
\end{align}
This series is convergent and the functions $f_{k}(h)$ satisfy 
\begin{align}
\lim\limits_{h\to 0}\frac{f_{k+1}(h)}{f_{k}(h)}=\lim\limits_{h\to 0}\frac{\widetilde{M}_{k+1}}{\widetilde{M}_{k}}\frac{h^{2(k+1)}}{h^{2k}}\frac{(2k)!}{(2k+2)!}\equiv 0
\end{align}
This means that each member of the set of filters approaches zero more rapidly than the previous member as $h\to 0$. Therefore the set of filters $\{\psi_{h,2k}\vert~k=0,1,2,...\}$ forms an asymptotic sequence in $h$. The difference between the true value $\varphi_{h}$ and approximate expression $\varphi_{h,n}$ goes to zero $(\varphi_{h}-\varphi_{h,n})/h^{n}\to 0$ as $h\to 0$. 

It is important to investigate how well an order-$n$ deconvolution filter $\varphi_{h,n}$ approximates the exact deconvolution filter $\varphi_{h}$ by comparing the moments. The exact moments of $L_{k}$ of $\varphi_{h}$ are readily computable without knowledge of the filter itself and are given by (\ref{deq:2017k8}). Based on the choice of $n$, the moments of the approximate deconvolution filter $\varphi_{h,n}$ are given by
\begin{align}
\text{exact}:\quad L_{k}&=\frac{M_k}{\Theta_k},
\quad\text{approx}:\quad L_{k}^{(n)}=\int_{V_h}\vert\vert\mathbf{r}^{\ast}\vert\vert^{2k+1}\varphi_{n}(\mathbf{r}^{\ast})d\vert\vert\mathbf{r}^{\ast}\vert\vert
\end{align}
Figure \ref{fig:momentCompute} indicates the moments of the deconvolution filter of the convolution filter given by (\ref{deq:2018d1}) with $p=5$. As can be clearly seen, the $k^{\text{th}}$ moment $L_{k}$ of an $n^{\text{th}}$  order approximate deconvolution filter $\varphi_{h,n}$ is approximately equal to the exact moment for $k=0,1,...,n$. Therefore, $\varphi_{h,n}$ approximates $\varphi_{h}$ to degree of accuracy $2n$. Formally,    
\begin{align}
L_{k}^{(n)}\simeq L_{k}\quad ^\forall k=0,1,2,\dots n \label{deq:2017k9}
\end{align}
Therefore, an $n^{\text{th}}$ order deconvolution filter cannot reproduce all moments since $\{L_{k}^{(n)}\neq L_{k}\vert~k=n+1,n+2,n+3,\dots\}$ will be inconsistent with the exact moments. 
However, as it will be shown below, exact deconvolution is unstable meaning that approximate filters (finite $n$) are more practical. 

\begin{figure}[H]
	\centering
	\begin{tabular}{|c|c|c|c|l|l|l|l|l|}
		\hline
		\multicolumn{1}{|>{\columncolor{kugray5}}c|}{Moment}&\multicolumn{4}{c|}{Approximate deconvolution filter}&\multirow{2}{*}{Exact $\varphi_{h}$}\\
		\arrayrulecolor{kugray5}
		\arrayrulecolor{black}
		\cline{2-5}
		\multicolumn{1}{|>{\columncolor{kugray5}}c|}{}&$\varphi_{h,1}$&$\varphi_{h,2}$&$\varphi_{h,3}$&$\varphi_{h,4}$&\\
		\hline
		$L_{0}$ &0.15915494&0.15915494&0.15915494&0.15915494&0.15915494\\
		$L_{1}$ &-0.0347461&-0.0347461&-0.0347469&-0.0347455&-0.0347461\\
		$L_{2}$ &-0.0457652$^\ast$&0.015423&0.0154208&0.015424&0.0154226\\
		$L_{3}$ &-0.0492201$^\ast$&0.0578682$^\ast$&-0.0104030&-0.010395&-0.0103981\\
		$L_{4}$ &-0.0580177$^\ast$&0.440901$^\ast$&-0.13588$^\ast$&0.00943111&0.00942152\\
		\hline
	\end{tabular}
	\caption[Sample kernel moments]{The first five moments of the approximate deconvolution filter for the convolution filter given by equation (\ref{deq:2018d1}) with $p=5$. The ($\ast$) indicates kernel inconsistency due to truncation.}
	\label{fig:momentCompute}
\end{figure}

\subsection{Convolution filters}
A natural choice for the convolution filter in SPH is the Gaussian filter \cite{Price2012}. In fact Monaghan \cite{Monaghan1992,Monaghan2005} suggests this as the first Golden Rule of SPH. All popular convolution filters are piecewise continuous polynomials including the B-spline functions \cite{Monaghan1985,Monaghan2005} and Wendland functions \cite{Wendland1995,Dehnen2012}. These filters are constructed to be Gaussian-like but they have compact support and give progressively better approximation to the Gaussian with higher filter order. 

For the purposes of the work presented here a new set of rational convolution filters are introduced as they have better interpolatory properties than their polynomial counterparts and their associated deconvolution filters are not singular. They are smoother and less oscillatory than polynomial filters. 
\begin{equation}
G_{h,p}= \left\{
\begin{array}{ll}
\alpha_{p}h^{-\nu}\left(\frac{1-\frac{1}{4h^{2}}\vert\vert\mathbf{r}-\mathbf{r}^{\prime}\vert\vert^{2}}{1+\frac{1}{4h^{2}}\vert\vert\mathbf{r}-\mathbf{r}^{\prime}\vert\vert^{2}}\right)^{p} & \vert\vert\mathbf{r}-\mathbf{r}^{\prime}\vert\vert\leq 2h \\

0 & \text{otherwise}\label{deq:2018d1}\\
\end{array} 
\right.
\end{equation}
It is also worth noting the relationship between $G_{h,p}$ and a gaussian filter
\begin{align}
G_{h,p}&\simeq\alpha_{p}h^{-\nu}\exp\left({-\pi\sqrt[\nu]{\alpha_{p}^{2}}\frac{\vert\vert\mathbf{r}-\mathbf{r}^{\prime}\vert\vert^{2}}{h^{2}}}\right)
\end{align}
The Weierstrass approximation theorem states that every continuous function defined on a closed interval $[\alpha, \beta]$ can be uniformly approximated as closely as desired by a polynomial function. 
\begin{mytheo}[Weierstrass approximation theorem]
	Suppose $G_{h}:\Omega_{h}(\mathbf{r})\to\mathbb{R}$ is a continuous real-valued function defined on the real, compact space $\Omega_{h}(\mathbf{r})$. For all $\mathbb{R}\ni\varepsilon>0$ there exists a polynomial $w_{h}:\Omega_{h}(\mathbf{r})\to\mathbb{R}$ such that for all $\mathbf{r},\mathbf{r}^{\prime}\in\Omega_{h}(\mathbf{r})$,we have,
	\begin{align}
	\vert G_{h}(\mathbf{r}-\mathbf{r}^{\prime})-w_{h}(\mathbf{r}-\mathbf{r}^{\prime})\vert<\varepsilon\quad\textup{or}\quad \vert\vert G_{h}(\mathbf{r}-\mathbf{r}^{\prime})-w_{h}(\mathbf{r}-\mathbf{r}^{\prime})\vert\vert<\varepsilon
	\end{align} 
\end{mytheo}
where $q:=\vert\vert\mathbf{r}-\mathbf{r}^{\prime}\vert\vert/h\in[-2,2]$

For the rational convolution filter (\ref{deq:2018d1}) with $p=7$, a possible approximating polynomial is given by
\begin{align}
w_{h}&=\left(1-\frac{q^2}{4}\right)^{8}\bigg(2048 - 3072q^{2} + 2816q^{4} - 1984q^{6}\nonumber\\
&\qquad+ 1184q^{8} - 628q^{10} + 305q^{12}\bigg)\label{deq:2018d2}
\end{align}
Later on, we will do spectral analysis in which the Fourier transforms of the filters will be required. The convolution filter (\ref{deq:2018d2}) is particularly useful in spectral analysis as it is difficult to compute the Fourier transform of (\ref{deq:2018d1}).

\def\PI{3.1415}
\def\LogFive{1.609437}
\def\ra{4096/(2*\PI*3*(108-5*\LogFive))}

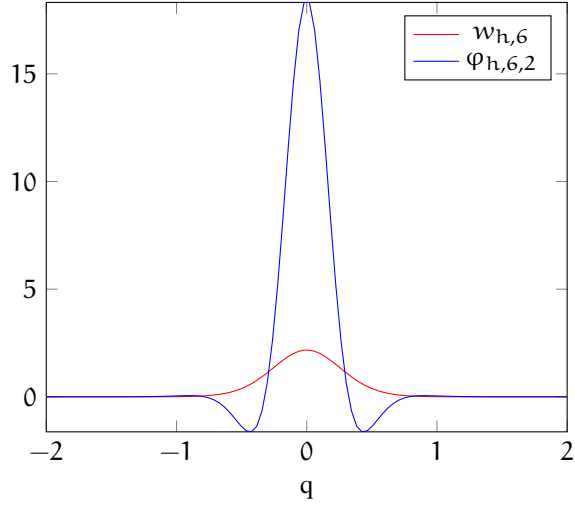
\begin{figure}[H]
	\centering
	\begin{tikzpicture}
	\begin{axis}[
	enlargelimits=false,
	xlabel = {q},
	%every axis x label/.style={
	%	at={(ticklabel* cs:1.05)},
	%	anchor=west,
	%},
	legend pos=north east
	]
	
	\addplot[red, domain=-2:2, samples=100]{\ra*(1-x^2/4)^6/(1+x^2)^6}node {};
	\addplot[blue, domain=-2:2, samples=100]{\ra*(1-x^2/4)^6/(1+x^2)^6+\ra*(15/2)*(1/(2*\PI*\ra))*(1-x^2/4)^4*(x^4-30*x^2+4)/(1+x^2)^8-\ra*(15/(4*2))*(1/(2*\PI*\ra))^2*(1-x^2/4)^2*(2*x^10-207*x^8+3844*x^6-18111*x^4+8808*x^2-528)/(1+x^2)^10}node {};
	\addlegendentry{$w_{h,6}$}
	\addlegendentry{$\varphi_{h,6,2}$}
	\end{axis}
	\end{tikzpicture} 
	\caption[Convolution filter and its associated deconvolution filter]{A plot of the convolution filter $w_{h,p}$ for p=6 (red) and the corresponding deconvolution filter $\varphi_{h,p,n}$ (blue) obtained by truncating the infinite sum at $n=2$. The filter cut-off length $h$ has been set to unity.}
	\label{graph:convDeconvfilters}
\end{figure}    

Note the sharpness and height of the deconvolution filter with respect to the convolution filter in figure \ref{graph:convDeconvfilters}. This means that the deconvolution filter approximates the Dirac delta function more accurately than the convolution filter. The other feature is that the deconvolution filter changes sign between positive and negative. This feature means that unlike the convolution filter, the deconvolution filter does not damp or attenuate high frequency components; this is important for the reconstructing the original continuum field.
 
\subsection{Spectral Analysis}
There are two factors that determine the upper limit of $n$; numerical stability and numerical instability. Increasing $n$ leads to improved accuracy and hence reduced numerical dissipation. Consider the Navier-Stokes equations, assuming Kolmogorov's -5/3 law, i.e. $E(k)\propto k^{-5/3}$, numerical dissipation can be approximated as a normalized coefficient

\begin{align}
c_{n,\text{dissp}} &= \frac{\int_{\mathbb{R}^{2}}\vert\vert\mathbf{k}\vert\vert^{2}E(\mathbf{k})d^{2}\mathbf{k}-\int_{\mathbb{R}^{2}}\vert\vert\mathbf{k}\vert\vert^{2}E(\mathbf{k})\vert\hat{\varphi}_{n}(\mathbf{k}h)\hat{w}(\mathbf{k}h)\vert^{2}d^{2}\mathbf{k}}{\int_{\mathbb{R}^{2}}\vert\vert\mathbf{k}\vert\vert^{2}E(k)d^{2}\mathbf{k}}\nonumber\\
&=\frac{\int_{0}^{k_{c}}k^{2}E(k)dk-\int_{0}^{k_{c}}k^{2}E(k)\vert\hat{\varphi}_{n}(kh)\hat{w}(kh)\vert^{2}dk}{\int_{0}^{k_{c}}k^{2}E(k)dk}\label{deq:2017j2}
\end{align}
where the dimensionless cut-off wavenumber ${k}^{\ast}_{c}:=k_{c}h=\pi$ is the cut-off wave number; the highest wavenumber that can be represented on a grid. The first term in the numerator is the exact dissipation and the second term is the restored dissipation, and thus the difference corresponds to numerical dissipation. 

on the other hand, numerical instability is much more difficult to quantify. Assuming the numerical error is due to spectral truncation, the energy error is on the order of $E_{e}(k)\propto k^{2}$ at high wavenumber. We can define a normalized numerical instability coefficient as
\begin{align}
c_{n,\text{instab}} &= \frac{\int_{\mathbb{R}^{2}}\vert\vert\mathbf{k}\vert\vert^{2}E_{e}(k)\vert\hat{\varphi}_{n}(\mathbf{k}h)\hat{w}(\mathbf{k}h)\vert^{2}d^{2}\mathbf{k}}{\int_{\mathbb{R}^{2}}\vert\vert\mathbf{k}\vert\vert^{2}E(\mathbf{k})d^{2}\mathbf{k}}\nonumber\\
&=\frac{\int_{0}^{k_{c}}k^{2}E_{e}(k)\vert\hat{\varphi}_{n}(kh)\hat{w}(kh)\vert^{2}dk}{\int_{0}^{k_{c}}k^{2}E_{e}(k)dk}\label{deq:2017j3}
\end{align}

\begin{myprop}
	Exact deconvolution has zero numerical dissipation.
	\begin{proof}
		We first compute the fourier transform of the completeness statement (\ref{deq:12}) to obtain
		\begin{align}
		\widehat{\varphi}(k^{\ast})\widehat{w}(k^{\ast})&=1\quad k^{\ast}:=kh\label{deq:2017j1}\\
		\therefore \lim\limits_{n\to\infty}c_{n,\text{dissp}}&=0\quad\text{by \ref{deq:2017j2}}
		\end{align}
		
	\end{proof}	
\end{myprop}
\begin{myprop}
	Exact deconvolution is numerically unstable.
	\begin{proof}
		From \ref{deq:2017j3} we have
		\begin{align}
		\lim\limits_{n\to\infty}c_{n,\text{instab}}&=1\quad\text{by \ref{deq:2017j3}}
		\end{align}
		
	\end{proof}	
\end{myprop}

Using Wendland's $C6$ filter as the convolution filter, by constructing its associated approximate deconvolution filter, the numerical dissipation and instability coefficients are plotted in the Fourier space as shown in figure \ref{fig:stability}. It can be noted that as the order of the deconvolution filter increases, numerical dissipation decreases as expected. However, as with decreased numerical dissipation there is nothing to mitigate any associated numerical instabilities hence the coefficient of numerical instability is large. The hypothesis used in \cite{Fang2017} that the energy error follows a square law i.e.  $E(\mathbf{k}^{\ast})\propto\vert\vert\mathbf{k}^{\ast}\vert\vert^{2}$ is qualitative and not necessarily accurate. There is further room for improvement in order to obtain a more accurate estimation of numerical instability.   
%%%%%%%%%%%%%%%%%%%%%%%%%%%%%%%%%%%%%%%%%%%
% compute numerical dissipation and numerical instbility for wendland c5 filter
%%%%%%%%%%%%%%%%%%%%%%%%%%%%%%%%%%%%%%%%%%%

\def\KA{9.574872826058547988067254487017211805215465622774582690406245985810575282862664567263901877421997828614334957770813734546463822047615749309828320074504326163648831735117856759778489292932119054339768999225590829622286012248344498301}
\def\KB{22.954620169272791494552124640272449183855846012507349495114000975244620990363708910246923836498829028474034381661298446867989975483484012415113504429553781557864959476399521654290591351750988834516319051322970621909951573388908165677}
\def\KC{32.168069877402718676213881335303495048074974514356714887102193212310999681973571973755447374540227620186840815135094949845453727907735880423586255458206000547347572121175402964139315930695417829320927151895008486012628903940389560828}
\def\KD{36.591796043303429871216698823479207787309535073638221577238066505111461623888333166611453424551835793573872425064420739185123481028836321381588049391781993763082106361769832353032252164239156951799412895788588803654201536583896358168}
\def\KN{38.924178248104418581876144306766420916930863830159496941393051801389549591286131013666895021312331642508816099090262915659968155075337281391394273164945650223627429922683780646009761695372789065053486912471240939410158049184219484961}

% numerical instability

\def\PA{99.12465873682033412234189139408927133971049071222938842210739828690024451081552441104840172651271280500483653547730306581902475950193796777021471833584010600035905259984003088772496364046622962541039328426023566240054043457086839761}
\def\PB{408.338002885996163780309933900729670956377603029772162949438320680129682252710019110055473457118065092154184008044776422090187516338202301119137526997188284997713482232131646807737332542289470139935016590283258047727792917624403940281}

\def\PC{719.060534670340225429340268247162439788102763485822728968679106881673591871256372528275955105771650780772980965635136647065477611174007256444386256440555782687469369561842600778320401754954778242249597340445862550029150885596840427646}
\def\PD{898.810688248942318347617835163582084793012442057124328211360093955424851017241257760739302833706160834825996907352317142986392246875499084795643154755450982940319008455178560946486459842590025328826417228510472236557829726552444112753}
\def\PN{1006.764409258930689171402164357347277305810825098467420918078164127711555764537293646150874363336578522878968290905363402829825128603736999670471570087856096614937617707097389158186679664625205204191576470383197339819252793391700320045}

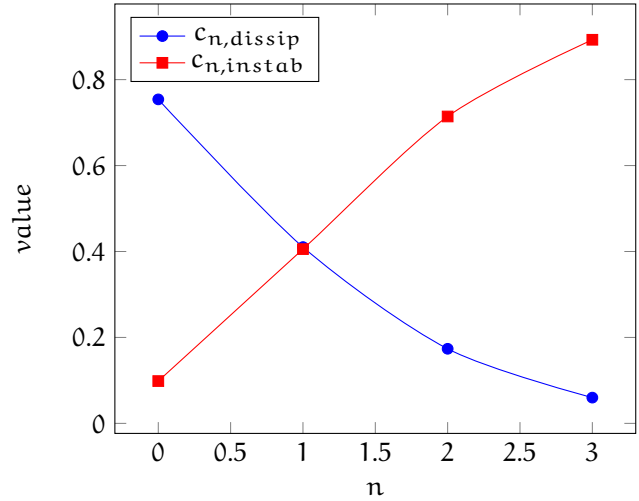
\begin{figure}[H]
	\begin{tikzpicture}
	\begin{axis}[
	xlabel=$n$,
	ylabel=$value$,legend pos = north west]
	\addplot[smooth,mark=*,blue] plot coordinates {
		(0,1-\KA/\KN)
		(1,1-\KB/\KN)
		(2,1-\KC/\KN)
		(3,1-\KD/\KN)
	};
	\addlegendentry{$c_{n,dissip}$}
	
	\addplot[smooth,color=red,mark=square*]
	plot coordinates {
		(0,\PA/\PN)
		(1,\PB/\PN)
		(2,\PC/\PN)
		(3,\PD/\PN)
	};
	\addlegendentry{$c_{n,instab}$}
	\end{axis}
	\end{tikzpicture}
	\caption[Numerical dissipation and instability for Wendland C5 filter]{Coefficients $c_{n,dissip}$ and $c_{n,instab}$ for the deconvolution filter associated with Wendland's C5 filter. Here $n$ correspond to the value at which the deconvolution filter $\varphi_{n}$ is truncated.}
	\label{fig:stability}
\end{figure}

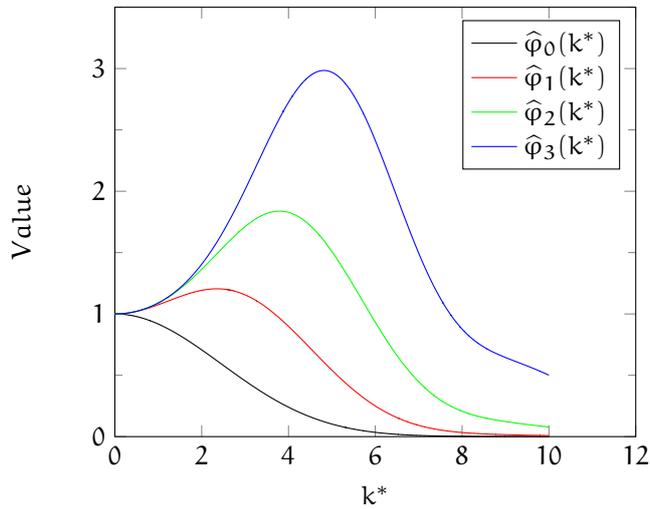
\begin{figure}[H]
	\begin{tikzpicture}
	\begin{axis}[
	%title = {Picture 1},  % whatever name you want
	xlabel = {$k^{\ast}$},
	ylabel = {$Value$},legend pos = north east,
	ymin = 0, ymax = 3.5,
	minor y tick num = 1,
	xmin = 0, xmax = 12,
	]
	\addplot[smooth, color=black] table {FTdFilter0.txt};
	\addlegendentry{$\widehat{\varphi}_{0}(k^{\ast})$}
	\addplot[smooth, color=red] table {FTdFilter1.txt};
	\addlegendentry{$\widehat{\varphi}_{1}(k^{\ast})$}
	\addplot[smooth, color=green] table {FTdFilter2.txt};
	\addlegendentry{$\widehat{\varphi}_{2}(k^{\ast})$}
	\addplot[smooth, color=blue] table {FTdFilter3.txt};
	\addlegendentry{$\widehat{\varphi}_{3}(k^{\ast})$}
	\end{axis}
	\end{tikzpicture}
	\caption[Fourier transform of  the deconvolution filter associated with Wendland $C6$ filter]{Fourier transform of the deconvolution filter $\{\widehat{\varphi}_{n}(k^{\ast})\vert n=0,1,2,3\}$ associated with Wendland $C6$ filter in physical space with $k^{\ast}=kh$ the non-dimensional wavenumber.}
	\label{fig:fourierDeconv}
\end{figure}

Figure \ref{fig:fourierDeconv} show a plot of the Fourier transform of the $n^{\text{th}}$ order deconvolution up to order $2n=6$. Note that at zeroth order, the deconvolution filter is identical to the convolution filter i.e. $w^{h}\equiv\varphi^{h}_{0}$ and the Fourier transform is shown by the solid black curve in Figure \ref{fig:fourierDeconv}. The range of wavenumbers over which the deconvolution filter has values greater than 1 increases with filter order. The implication of this characteristic shape that the deconvolution filter in Fourier space is twofold. First, the deconvolution filter is able to restore the low frequency components, consistent with its mathematical property that it becomes a Dirac delta function in the limit $h\to 0$; see equation (\ref{deq:2017e6}) for a proof of this. Second, the deconvolution filter damping the high frequency components for stability reasons. It will be shown that exact deconvolution is unstable, thus approximate deconvolution which does not damp high frequency components is what is practically usable \cite{Fang2017}. 

\begin{figure}[H]
	\begin{tikzpicture}
	\begin{axis}[
	%title = {Picture 1},  % whatever name you want
	xlabel = {$k^{\ast}$},
	ylabel = {$Value$},legend pos = north east,
	ymin = 0, ymax = 1.2,
	minor y tick num = 1,
	xmin = 0, xmax = 10,
	]
	\addplot[smooth, color=black] table {PdFilter0.txt};
	\addlegendentry{$\widehat{\varphi}_{0}(k^{\ast})\widehat{w}(k^{\ast})$}
	\addplot[smooth, color=red] table {PdFilter1.txt};
	\addlegendentry{$\widehat{\varphi}_{1}(k^{\ast})\widehat{w}(k^{\ast})$}
	\addplot[smooth, color=green] table {PdFilter2.txt};
	\addlegendentry{$\widehat{\varphi}_{2}(k^{\ast})\widehat{w}(k^{\ast})$}
	\addplot[smooth, color=blue] table {PdFilter3.txt};
	\addlegendentry{$\widehat{\varphi}_{3}(k^{\ast})\widehat{w}(k^{\ast})$}
	\end{axis}
	\end{tikzpicture}
	\caption[Fourier spectrum of the approximate completeness statement $\widehat{\varphi}_{2}(k^{\ast})\widehat{w}(k^{\ast})$]{Fourier transform of the deconvolution filter $\{\widehat{\varphi}_{n}(k^{\ast})\widehat{w}(k^{\ast})\vert~ n=0,1,2,3\}$ associated with Wendland $C6$ filter in physical space with $k^{\ast}=kh$ the non-dimensional wavenumber.}
	\label{fig:spectrumADM}
\end{figure}
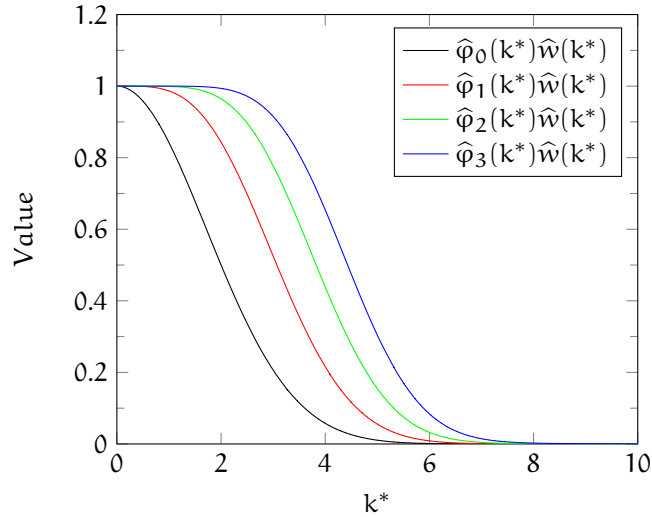

The next thing worth investigating is the resolution of identity given by equation (\ref{deq:12}). In Fourier space, we have the spectrum $\hat{\varphi}(\mathbf{k}^{\ast})\hat{w}(\mathbf{k}^{\ast})=1$ for all normalized wavenumbers $\mathbf{k}^{\ast}$. For the ADM, the resolution of identity is not exact and has a spectrum given by $\hat{\varphi}_{n}(\mathbf{k}^{\ast})\hat{w}(\mathbf{k}^{\ast})$ whose plot in the Fourier space is shown in figure \ref{fig:spectrumADM}. It shows that the higher the order of the deconvolution filter, the larger the bandwidth of wavenumbers restored. We also see that in the limit $n\to\infty$, we have $\hat{\varphi}_{n}(\mathbf{k}^{\ast})\hat{w}(\mathbf{k}^{\ast})=1$ as expected from the theory presented in here.

\section{Conclusion}
In this paper a new method called unsmoothed particle hydrodynamics SPH$-i$ has been proposed as a complete form of Smoothed Particle Hydrodynamics. 
In order to improve the performance and mathematical consistency of SPH, a transform pair called FIT and DIT has been proposed. First a version of SPH that is consistent with explicit LES was derived \cite{Cholas2018} using the FIT.  For completeness, and hence to avoid the problem of turbulence modeling, the filtered equations where then de-filtered using the DIT leading to a new model called SPH$-i$. 

A rigorous procedure for deriving convolution and deconvolution operators from the transform pair has been given. Using these operators, a method for constructing an exact, compatible and unique deconvolution filter has been proposed. Following the discussion that this exact deconvolution filter is unstable, an approximate deconvolution filter has been proposed; the ideas of ADM are studied in many areas of science such as LES and image processing \cite{San2011}\cite{Sandri1990}\cite{Domingo2015}\cite{Germano2009}\cite{Ulmer2010} \cite{Geurts1997} \cite{Masry1992} \cite{Wang2017} and convergence of ADMs \cite{Berselli2012}.

Another major difference between standard SPH and the proposed SPH$-i$ is that writing the correct Lagrangian using the smoothed variables in SPH is actually non-trivial whereas the one for SPH$-i$ is easily given by Eckart's Lagrangian \cite{Eckart1960}.  

Finally, being a high order model, SPH$-i$ should in principle be more accurate than standard SPH.

\renewcommand{\refname}{\spacedlowsmallcaps{References}} % For modifying the bibliography heading

\bibliographystyle{unsrt}

\bibliography{articlez} % The file containing the bibliography

%----------------------------------------------------------------------------------------
\clearpage
\appendixheaderon
\begin{appendices}
	\section{The SPH-$\lowercase{i}$ Laplacian}\label{ap:a}
	Here we present the procedure for constructing the Laplacian operator in the SPH$-i$ model. First we note the following identities
	\begin{align}
	\nabla^{2}(\kappa p)&=\kappa\nabla^{2}p +2\bm{\nabla}\kappa\cdot\bm{\nabla} p +p\nabla^{2}\kappa\label{app:a1}\\
	\bm{\nabla}\cdot(\kappa\bm{\nabla}p)&=\kappa\nabla^{2}p +\bm{\nabla}\kappa\cdot\bm{\nabla} p\label{app:aa1}
	\end{align}
	Next, we define the following integral identities
	\begin{align}
	\mathcal{J}_{1}&\stackrel{\text{def}}{=}\int_{\Omega}\bigg(\langle\kappa_{h}(\mathbf{r}^{\prime})\rangle\bm{\nabla}p(\mathbf{r})\cdot\bm{\nabla}\varphi_{h}+\kappa(\mathbf{r})\bm{\nabla}^{\prime}\langle p_{h}(\mathbf{r}^{\prime})\rangle\cdot\bm{\nabla}\varphi_{h}\bigg)d^{\nu}\mathbf{r}^{\prime}\label{app:a3}\\
	&=\bm{\nabla}p\cdot\bm{\nabla}\int_{\Omega}\langle\kappa_{h}(\mathbf{r}^{\prime})\rangle\varphi_{h}d^{\nu}(\mathbf{r}^{\prime})+\kappa\bm{\nabla}\cdot\int_{\Omega}\bm{\nabla}^{\prime}\langle p_{h}(\mathbf{r}^{\prime})\rangle\cdot\varphi_{h}d^{\nu}\mathbf{r}^{\prime}\nonumber\\
	&\equiv\bm{\nabla}p\cdot\bm{\nabla}\kappa+\kappa\nabla^{2}p\qquad\text{by the DIT}\label{app:a4}\\
	\mathcal{J}_{2}&\stackrel{\text{def}}{=}\int_{\Omega}\bigg(\langle p_{h}(\mathbf{r}^{\prime})\rangle\bm{\nabla}\kappa(\mathbf{r})\cdot\bm{\nabla}\varphi_{h}+p(\mathbf{r})\bm{\nabla}^{\prime}\langle \kappa_{h}(\mathbf{r}^{\prime})\rangle\cdot\bm{\nabla}\varphi_{h}\bigg)d^{\nu}\mathbf{r}^{\prime}\label{app:a5}\\
	&=\bm{\nabla}\kappa\cdot\bm{\nabla}\int_{\Omega}\langle p_{h}(\mathbf{r}^{\prime})\varphi_{h}d^{\nu}(\mathbf{r}^{\prime})+p\bm{\nabla}\cdot\int_{\Omega}\bm{\nabla}^{\prime}\langle \kappa_{h}(\mathbf{r}^{\prime})\rangle\cdot\varphi_{h}d^{\nu}\mathbf{r}^{\prime}\nonumber\\
	&\equiv\bm{\nabla}\kappa\cdot\bm{\nabla}p+p\nabla^{2}\kappa\qquad\text{by the DIT}\label{app:a6}
	\end{align} 
	If we now add the two identities (\ref{app:a4}) and (\ref{app:a6}) together with (\ref{app:a1}) we obtain
	\begin{align}
	\mathcal{J}_{1}+\mathcal{J}_{2}&=\nabla^{2}(p\kappa)\label{app:a2}
	\end{align}
	The next step involves the finite difference approximation of $\mathcal{J}_{1}$. By the Taylor expansion to get the following
	\begin{align}
	\delta\mathbf{r}\cdot\bm{\nabla}p(\mathbf{r})=p(\mathbf{r})-p(\mathbf{r}^{\prime})+\mathcal{O}\left(\vert\vert\delta\mathbf{r}\vert\vert^{2}\right)\nonumber\\
	\delta\mathbf{r}\cdot\bm{\nabla}^{\prime}\langle p(\mathbf{r}^{\prime})\rangle=\langle p(\mathbf{r})\rangle-\langle p_{h}(\mathbf{r}^{\prime})\rangle+\mathcal{O}\left(\vert\vert\delta\mathbf{r}\vert\vert^{2}\right)\label{app:a7}
	\end{align}
	where $\delta\mathbf{r}:=\mathbf{r}-\mathbf{r}^{\prime}$ is the relative position between the two interacting particles. Thus direct substitution of (\ref{app:a7}) into (\ref{app:a3}) yields
	\begin{align}
	\mathcal{J}_{1}&=\int_{\Omega}\bigg\{\langle\kappa_{h}(\mathbf{r}^{\prime})\rangle\bigg(p(\mathbf{r})-p(\mathbf{r}^{\prime})\bigg)+\kappa(\mathbf{r})\bigg(\langle p_{h}(\mathbf{r})\rangle-\langle p_{h}(\mathbf{r}^{\prime})\rangle\bigg)\bigg\}\frac{\delta\mathbf{r}\cdot\bm{\nabla}\varphi_{h}}{\vert\vert\delta\mathbf{r}\vert\vert^{2}}d^{\nu}\mathbf{r}^{\prime}\label{app:a8}
	\end{align}
	Following this same procedure, the finite difference approximation of $\mathcal{J}_{2}$ becomes
	\begin{align}
	\mathcal{J}_{2}&=\int_{\Omega}\bigg\{\langle p_{h}(\mathbf{r}^{\prime})\rangle\bigg(\kappa(\mathbf{r})-\kappa(\mathbf{r}^{\prime})\bigg)+p(\mathbf{r})\bigg(\langle \kappa_{h}(\mathbf{r})\rangle-\langle \kappa_{h}(\mathbf{r}^{\prime})\rangle\bigg)\bigg\}\frac{\delta\mathbf{r}\cdot\bm{\nabla}\varphi_{h}}{\vert\vert\delta\mathbf{r}\vert\vert^{2}}d^{\nu}\mathbf{r}^{\prime}\label{app:a9}
	\end{align}
	Plugging (\ref{app:a8}) and (\ref{app:a9}) into (\ref{app:a2}) yields the following identity
	\begin{align}
	\nabla^{2}(p\kappa)&=\int_{\Omega}\bigg\{\langle\kappa_{h}(\mathbf{r}^{\prime})\rangle\bigg(p(\mathbf{r})-p(\mathbf{r}^{\prime})\bigg)+\kappa(\mathbf{r})\bigg(\langle p_{h}(\mathbf{r})\rangle-\langle p_{h}(\mathbf{r}^{\prime})\rangle\bigg)\nonumber\\
	&+\langle p_{h}(\mathbf{r}^{\prime})\rangle\bigg(\kappa(\mathbf{r})-\kappa(\mathbf{r}^{\prime})\bigg)+p(\mathbf{r})\bigg(\langle \kappa_{h}(\mathbf{r})\rangle-\langle \kappa_{h}(\mathbf{r}^{\prime})\rangle\bigg)\bigg\}\frac{\delta\mathbf{r}\cdot\bm{\nabla}\varphi_{h}}{\vert\vert\delta\mathbf{r}\vert\vert^{2}}d^{\nu}\mathbf{r}^{\prime}\label{app:a10}
	\end{align}
	Furthermore, two more identities can now be extracted from (\ref{app:a10}) i.e. when either $\kappa$ or $p$ is constant.
	\begin{align}
	\kappa\nabla^{2}p&\simeq\int_{\Omega}\bigg\{\langle\kappa_{h}(\mathbf{r})\rangle\bigg(p(\mathbf{r})-p(\mathbf{r}^{\prime})\bigg)+\kappa(\mathbf{r})\bigg(\langle p_{h}(\mathbf{r})\rangle-\langle p_{h}(\mathbf{r}^{\prime})\rangle\bigg)\bigg\}\frac{\delta\mathbf{r}\cdot\bm{\nabla}\varphi_{h}}{\vert\vert\delta\mathbf{r}\vert\vert^{2}}d^{\nu}\mathbf{r}^{\prime}\label{app:a11}\\
	p\nabla^{2}\kappa&\simeq\int_{\Omega}\bigg\{\langle p_{h}(\mathbf{r})\rangle\bigg(\kappa(\mathbf{r})-\kappa(\mathbf{r}^{\prime})\bigg)+p(\mathbf{r})\bigg(\langle \kappa_{h}(\mathbf{r})\rangle-\langle \kappa_{h}(\mathbf{r}^{\prime})\rangle\bigg)\bigg\}\frac{\delta\mathbf{r}\cdot\bm{\nabla}\varphi_{h}}{\vert\vert\delta\mathbf{r}\vert\vert^{2}}d^{\nu}\mathbf{r}^{\prime}\label{app:a12}
	\end{align}
	We get another identity by substituting (\ref{app:a10}), (\ref{app:a11}) and (\ref{app:a12}) into (\ref{app:a1}) we obtain
	\begin{align}
	\bm{\nabla}\kappa\cdot\bm{\nabla} p&=-\frac{1}{2}\int_{\Omega}\bigg\{\bigg(\langle\kappa_{h}(\mathbf{r})\rangle-\langle\kappa_{h}(\mathbf{r}^{\prime})\rangle\bigg)\bigg(p(\mathbf{r})-p(\mathbf{r}^{\prime})\bigg)\nonumber\\
	&\qquad+\bigg(\kappa(\mathbf{r})-\kappa(\mathbf{r}^{\prime})\bigg)\bigg(\langle p_{h}(\mathbf{r})\rangle-\langle p_{h}(\mathbf{r}^{\prime})\rangle\bigg)\bigg\}\frac{\delta\mathbf{r}\cdot\bm{\nabla}\varphi_{h}}{\vert\vert\delta\mathbf{r}\vert\vert^{2}}d^{\nu}\mathbf{r}^{\prime}\label{app:a13}
	\end{align} 
	Finally, substitution of (\ref{app:a11}) and (\ref{app:a13}) into (\ref{app:aa1}) yields the desired result.
	\begin{align}
	\bm{\nabla}\cdot(\kappa\bm{\nabla}p)&=\frac{1}{2}\int_{\Omega}\bigg\{\bigg(\langle\kappa_{h}(\mathbf{r})\rangle+\langle\kappa_{h}(\mathbf{r}^{\prime})\rangle\bigg)\bigg(p(\mathbf{r})-p(\mathbf{r}^{\prime})\bigg)\nonumber\\
	&\qquad+\bigg(\kappa(\mathbf{r})+\kappa(\mathbf{r}^{\prime})\bigg)\bigg(\langle p_{h}(\mathbf{r})\rangle-\langle p_{h}(\mathbf{r}^{\prime})\rangle\bigg)\bigg\}\frac{\delta\mathbf{r}\cdot\bm{\nabla}\varphi_{h}}{\vert\vert\delta\mathbf{r}\vert\vert^{2}}d^{\nu}\mathbf{r}^{\prime}\label{app:a14}
	\end{align} 
\end{appendices}

\end{document}